\documentclass[twocolumn,showpacs,aps,prx,amsmath,amssymb,superscriptaddress]{revtex4-1}

\usepackage{graphicx}
\usepackage{dcolumn}
\usepackage{bm}
\usepackage{hyperref}
\usepackage{subfigure}

\usepackage{epstopdf}
\usepackage{tikz}

\usepackage{color,bbm}

\newcommand{\beq}{\begin{equation}}
\newcommand{\eeq}{\end{equation}}
\newcommand{\bqa}{\begin{eqnarray}}
\newcommand{\eqa}{\end{eqnarray}}
\newcommand{\nn}{\nonumber}

\newcommand{\rt}[1]{\sqrt{#1}\,}

\newcommand{\bra}[1]{ \langle{#1} |}
\newcommand{\ket}[1]{ |{#1} \rangle}

\newcommand{\sq}[1]{\left[ {#1} \right]}

\newcommand{\an}[1]{\left\langle{#1}\right\rangle}
\newcommand{\tr}[1]{{\rm Tr}\sq{ {#1} }}

\newcommand{\id}{\mathbbm{1}}

\newtheorem{Theorem}{Theorem}

\newtheorem{Corollary}{Corollary}
\newtheorem{Lemma}{Lemma}
\newtheorem{Proposition}{Proposition}

\definecolor{maroon}{rgb}{0.7,0,0}

\definecolor{ngreen}{rgb}{0.3,0.7,0.3}

\definecolor{golden}{rgb}{0.8,0.6,0.1}

\begin{document}

\title{Faithful geometric measures for genuine tripartite entanglement}

\author{Xiaozhen Ge}
\affiliation{The Department of Control Science and Engineering, Tongji University, Shanghai 201804, China}
\affiliation{Department of Applied Mathematics, The Hong Kong Polytechnic University, Hong Kong, China}

\author{Lijun Liu}
\email{lljcelia@126.com}
\affiliation{Department of Mathematics and Computer Science, Shanxi Normal University, Taiyuan 030006, China}

\author{Yong Wang}
\affiliation{The Department of Control Science and Engineering, Tongji University, Shanghai 201804, China}

\author{Yu Xiang}

\affiliation{State Key Laboratory for Mesoscopic Physics, School of Physics, Frontiers Science Center for Nano-optoelectronics, $\&$ Collaborative Innovation Center of Quantum Matter, Peking University, Beijing 100871, China}

\author{Guofeng Zhang}
\affiliation{Department of Applied Mathematics, The Hong Kong Polytechnic University, Hong Kong, China}

\author{Li Li}
\affiliation{The Department of Control Science and Engineering, Tongji University, Shanghai 201804, China}

\author{Shuming Cheng}
\email{drshuming.cheng@gmail.com}
\affiliation{The Department of Control Science and Engineering, Tongji University, Shanghai 201804, China}
\affiliation{Shanghai Institute of Intelligent Science and Technology, Tongji University, Shanghai 201804, China}
\affiliation{Institute for Advanced Study, Tongji University, Shanghai, 200092, China}

\date{\today}

\begin{abstract}
	
 We present a faithful geometric picture for genuine tripartite entanglement of discrete, continuous, and hybrid quantum systems. We first show that the triangle relation $\mathcal{E}^\alpha_{i|jk}\leq \mathcal{E}^\alpha_{j|ik}+\mathcal{E}^\alpha_{k|ij}$ holds for all subadditive bipartite entanglement measure $\mathcal{E}$, all permutations under parties $i, j, k$, all $\alpha \in [0, 1]$, and all pure tripartite states. It provides a geometric interpretation that bipartition entanglement, measured by $\mathcal{E}^\alpha$, corresponds to the side of a triangle, of which the area with $\alpha \in (0, 1)$ is nonzero if and only if the underlying state is genuinely entangled. Then, we rigorously prove the non-obtuse triangle area with $0<\alpha\leq 1/2$ is a measure for genuine tripartite entanglement. Useful lower and upper bounds for these measures are obtained, and generalisations of our results are also presented. Finally, it is significantly strengthened for qubits that, given a set of subadditive and non-additive measures, some state is always found to violate the triangle relation for any $\alpha>1$, and the triangle area is not a measure for any $\alpha>1/2$. Hence, our results are expected to aid significant progress in studying both discrete and continuous multipartite entanglement. 
 
\end{abstract}


\maketitle

\section{Introduction}

 Entanglement is the most puzzling feature of quantum theory, and plays an indispensable role in various quantum processing tasks, such as device-independent cryptography~\cite{Ekert1991} and teleportation~\cite{Bennett1993}. It is thus of fundamental and practical interest to investigate the characterisation and quantification of entanglement. During past decades, extensive research are devoted to studying entanglement for bipartite quantum systems, and correspondingly, numerous methods have been developed to quantify bipartite entanglement~\cite{Horodecki2009,Guhne2009,Weedbrook2012,Plenio2014,Adesso2014}. However, much less progress has been achieved for multipartite entanglement, mainly due to complicated structures of multipartite states and operations~\cite{Chitambar2014}. Besides, it still lacks an unified approach to quantifying entanglement of both discrete and continuous quantum systems, except for using a well-chosen entanglement witness~\cite{Guhne2009} or Fisher information matrix~\cite{Gessner2016} to certify the presence of entanglement.

In this work, we mainly study the entanglement properties of discrete, continuous, and even hybrid tripartite systems. In particular, using the well-explored measures for bipartite entanglement, we present an unified geometric approach to tackle the problems of: 1) how to characterise tripartite entanglement and 2) how to quantify genuine tripartite entanglement via {\it proper} measures that are able to detect all genuinely entangled states useful in multiparty information tasks.

Our first main result is a geometric characterisation of tripartite entanglement based on a class of triangle relations. Specifically, given any tripartite state $\ket{\psi}_{ABC}$, the degree of entanglement between one party and the rest, quantified by a subadditive measure $\mathcal{E}$, satisfies
 \beq\label{triangle}
\mathcal{E}_{i|jk}^\alpha\leq \mathcal{E}^\alpha_{j|ik}+\mathcal{E}^\alpha_{k|ij}
\eeq
and its permutations under three parties $i, j, k =A, B, C$ for power $\alpha\in [0,1]$. This relation suggests that entanglement owned by one party is no larger than the sum of entanglement by the other two, complementary to monogamy of entanglement~\cite{Coffman2000,Dhar2017} and closely relate to the entanglement polytopes~\cite{Higuchi2003,Walter2013} and the quantum marginal problem~\cite{Klyachko2006,Eisert2008}. Furthermore, Tsallis-$2$ entropy~\cite{Tsallis1988} is chosen as measure $\mathcal{E}$ to exemplify that it is valid for all discrete, all Gaussian, and all discrete-discrete-continuous pure tripartite states, significantly generalising previous results for qubits~\cite{Zhu2015,Qian2018, Xie2023,Beckey2021}. When it comes to qubit, Eq.~(\ref{triangle}) is obtained for subadditive measures, such as von Neumann entropy~\cite{Nielsen2000}, Tsallis entropy~\cite{Tsallis1988,Hu2006}, squared concurrence~\cite{Wootters1998,Rungta2001,Albeverio2001}, squared negativity~\cite{Vidal2002}, and non-subadditive ones, including Schmidt weight~\cite{Grobe1994,Qi2018} and R\'enyi-$2$ entropy~\cite{Renyi1961}.

\begin{figure}
 	\begin{center}
 		\includegraphics[width= 1.0\columnwidth]{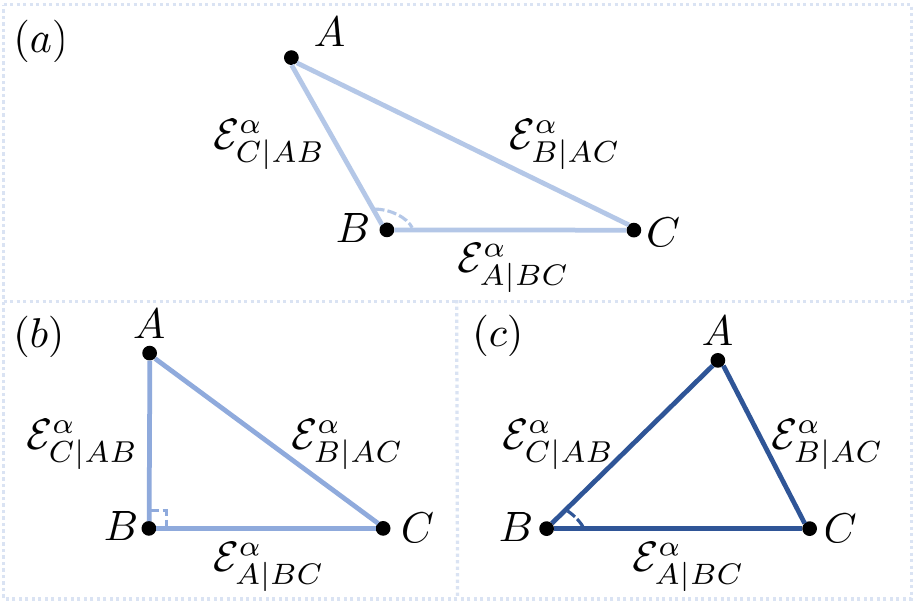}
 	\end{center}
 	\caption{The faithful geometric interpretation for genuine tripartite entanglement of discrete, continuous, and hybrid quantum systems. Following the triangle relation~(\ref{triangle}) for any pure state $\ket{\psi}_{ijk}$ shared by three parties $i, j, k=A, B, C$, three bipartition entanglement, measured by $\mathcal{E}^\alpha_{i|jk}$ with $\alpha\in [0,1]$, can be interpreted as the side of a triangle. The faithfulness is obtained in Proposition~\ref{proposition2} for generic tripartite states and Theorem~\ref{theorem2} for three-qubit states that the triangle has nonzero area if and only if the state is genuinely entangled. Depending on whether the condition~(\ref{nonobtuse}) is satisfied, the triangle can be categorised into: (a) obtuse, e.g.  Tsallis entropy~(\ref{Tsallis}) with $\alpha>1/2$ for some state; (b) right-angled, e.g., Schmidt weight with $\alpha=1/2$ for a class of W-class states; and (c) acute, e.g., all subadditive measures with $\alpha\in[0,1/2)$ for all pure states. In the last two cases, the triangle area~(\ref{area}) satisfies LOCC-monotonicity as per Theorems~\ref{theorem1} and~\ref{theorem3}.    }
 	\label{trianglefig}
 \end{figure}

As illustrated in Fig.~\ref{trianglefig}, the triangle relation~(\ref{triangle}) and its permutations provide us a nice geometric picture for tripartite entanglement, in the sense that its bipartition entanglement, measured by $\mathcal{E}^\alpha$, can be interpreted as the side of a triangle. It is further proven to be {\it faithful} for genuine tripartite entanglement that the induced triangle with $\alpha \in (0, 1)$ is nondegenerate, or equivalently, has nonzero area, if and only if the tripartite state, either pure or mixed, is genuinely entangled.

Our second main result is a class of faithful measures for genuine tripartite entanglement. Particularly, the triangle area, induced by the above relation~(\ref{triangle})
\begin{align}\label{area}
	\mathcal{A}(|\psi\rangle_{ijk})=\sqrt{Q(Q-\mathcal{E}^\alpha_{i|jk})(Q-\mathcal{E}^\alpha_{j|ik})(Q-\mathcal{E}^\alpha_{k|ij})}
\end{align}
with the semiperimeter $Q=(\mathcal{E}^\alpha_{i|jk}+\mathcal{E}^\alpha_{j|ik}+\mathcal{E}^\alpha_{k|ij})/2$, is a natural quantifier for genuine tripartite entanglement. We analytically prove that for any subadditive measure $\mathcal{E}$ with $\alpha\in (0, 1/2]$, the area is monotonic under local operations and classical communication (LOCC), thus being a reliable entanglement measure. Importantly, since the proof of LOCC-monotonicity is independent of measure and state, it is widely applicable to the discrete and/or continuous systems. Useful lower and upper bounds are derived for these geometric measures, related to the well-known multipartite entanglement measures, such as genuinely multipartite concurrence~\cite{Ma2011} and global entanglement measure~\cite{Meyer2002,Brennen2003}.

 Finally, our results are significantly strengthened for qubits, in the sense that, given a set of entanglement measures, some state is always found to violate the triangle relation~(\ref{triangle}) with any $\alpha>1$ and to violate the LOCC-monotonicity of triangle area~(\ref{area}) with any $\alpha>1/2$. As byproduct, our results confirm the concurrence fill~\cite{Ge2023,Jin2023} and the ergotropic fill~\cite{Puliyil2022} as feasible entanglement measures, and overcome an incompleteness in the proof in~\cite{Jin2023} to show the LOCC-monotonicity of the concurrence area. 

This work is organized as follows. In Sec.~\ref{entanglement}, we present a brief introduction to entanglement measures and the subadditivity. Then, we show that the triangle relation~(\ref{triangle}) holds for all subadditive measures and all pure tripartite states and provides a nice geometric picture for genuine tripartite entanglement in Sec.~\ref{trianglerelation}. In Sec.~\ref{areameasure}, we analytically prove that the triangle area~(\ref{area}) with $\alpha\in (0, 1/2]$ is a faithful measure for genuine tripartite entanglement, which is further exemplified via Tsallis-2 entropy to show its universal validity. We derive some useful upper and lower bounds for the triangle area in Sec.~\ref{bounds} and significantly strengthen our results for qubits in Sec.~\ref{qubits}. Finally, Sec.~\ref{discussions} concludes with discussions and possible generalizations.

\section{Entanglement measures}\label{entanglement}

The resource theory of entanglement~\cite{Vidal2000,Chitambar2019} is first briefly recapped. Within this framework, non-entangled states correspond to free states, while entangled ones can be recognised as essential resources to accomplish impossible tasks in the classical realm. In order to quantify these resources, an entanglement measure $\mathcal{E}$ is typically introduced as some function which maps any quantum state to a nonnegative number i.e., $\mathcal{E}(\rho)\geq 0$ for state $\rho$. Further, it needs to meet extra requirements~\cite{Vedral1997,Horodecki2009,Guhne2009,Szalay2015}: 1) faithfulness: $\mathcal{E}(\rho)=0$ if and only if $\rho$ is separable or non-entangled; 2) LOCC-monotonicity: $\mathcal{E}(\rho) \geq \sum_m p_m\,\mathcal{E}(\rho_m)$ for any $\rho$ and its LOCC-ensemble $\{p_m, \rho_m\}$, requiring entanglement never increases under free operations of LOCC; 3) symmetry: given an pure bipartite state $\ket{\psi}_{ij}$, global entanglement is determined by the local state, i.e., $\mathcal{E}(\ket{\psi}_{ij})\equiv E(\rho_i)= E(\rho_j)$, where $E$ is properly defined on states $\rho_{i(j)}={\rm Tr}_{j(i)}(\ket{\psi}_{ij}\bra{\psi})$. 

 One notable example satisfying the above conditions is Tsallis entropy~\cite{Tsallis1988}
\beq
\mathcal{T}(\ket{\psi}_{ij})= \mathcal{T}(\rho_i)\equiv\frac{1-{\rm Tr}(\rho_i^q)}{q-1},~~q\geq1 \label{Tsallis}
\eeq
for any pure bipartite state $\ket{\psi}_{ij}$. It recovers von Neumann entropy in the limit $q\rightarrow 1$ and reduces to linear entropy or generalised concurrence~\cite{Rungta2001,Albeverio2001} by $q=2$, both of which also admit the subadditivity~\cite{Auden2007,Araki1970}
\beq
E(\rho_{ij})\leq E(\rho_i)+E(\rho_j)
\eeq
 for any $\rho_{ij}$ in discrete and continuous systems. Indeed, whether all these requirements can be satisfied depends on both the state space and the measure, and there exist measures without subadditivity~\cite{Li2014}. For example, the measure of R\'enyi-$2$ entropy is not subadditive for qudits~\cite{Linden2013}, while it is for the Gaussian~\cite{Adesso2012}. In the following, we discuss how to use bipartite entanglement measures to study multipartite entanglement.

\section{Triangle relations and geometric picture for tripartite entanglement}\label{trianglerelation}

Any pure tripartite state $\ket{\psi}_{ijk}$ admits three bipartition among parties $i, j, k$, of which bipartite entanglement is quantified by $\mathcal{E}_{i|jk}, \mathcal{E}_{j|ik}, \mathcal{E}_{k|ij}$ respectively, with a bipartite entanglement measure $\mathcal{E}$. We can obtain the following result.

\begin{Proposition}	\label{proposition1}
For any subadditive measure $\mathcal{E}$, the triangle relation~(\ref{triangle}) holds for all pure tripartite states, all permutations under three parties, and all $\alpha \in [0, 1]$. 
\end{Proposition}

The proof is as follows. First, note from the symmetric property that $\mathcal{E}_{k|ij}\equiv E(\rho_{ij})=E(\rho_k)$ holds, with $\rho_{ij(k)}={\rm Tr}_{k(ij)}(\ket{\psi}_{ijk}\bra{\psi})$. It then follows from the subadditivity that $\mathcal{E}_{k|ij}=E(\rho_{ij})\leq E(\rho_i)+E(\rho_j)=\mathcal{E}_{i|kj}+\mathcal{E}_{j|ki}$, proving Eq.~(\ref{triangle}) with $\alpha=1$. Finally, we have  
\begin{eqnarray}
	\mathcal{E}^\alpha_{i|jk}\leq (\mathcal{E}_{j|ik}+\mathcal{E}_{k|ij})^\alpha \leq \mathcal{E}^\alpha_{j|ik}+\mathcal{E}^\alpha_{k|ij}, \forall~\alpha \in [0, 1). \label{leq1}
\end{eqnarray}
The first inequality follows from $x^r\leq y^r$ for nonnegative $x\leq y$ and $r<1$, and the second from $(x+y)^r\leq x^r+y^r$ for nonnegative $x,y$ and $r<1$. Thus, the triangle relation~(\ref{triangle}) holds for all pure states and all $\alpha\in[0,1]$, and its permutations can be obtained similarly. 


As illustrated in Fig.~\ref{trianglefig}, the triangle relation~(\ref{triangle}) yields a geometric description of $\ket{\psi}_{ijk}$ that bipartite entanglement corresponds to the side of a triangle. If the state is biseparable, i.e., at least one zero $\mathcal{E}^\alpha_{i|jk}$, then its triangle degenerates to a line or a point. Next, we show the converse is also true.

\begin{figure}
	\begin{center}
		\includegraphics[width= 1.0\columnwidth]{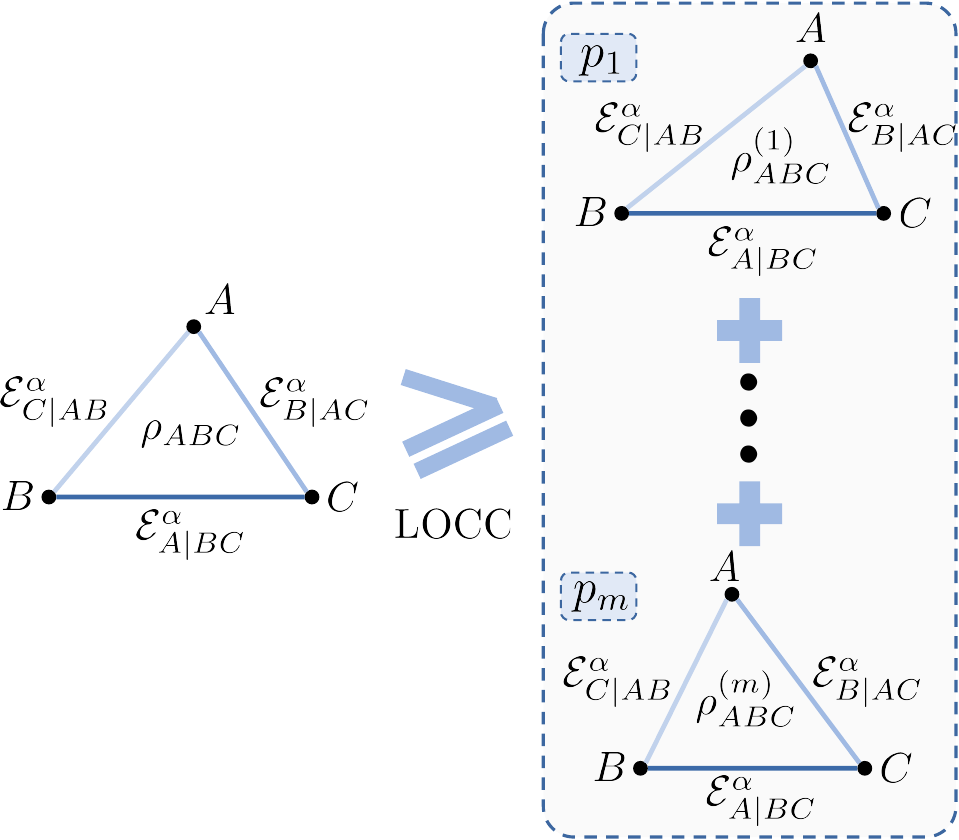}
	\end{center}
	\caption{LOCC-monotonicity of the triangle area~(\ref{area}) with $\alpha\in(0,1/2]$, which is proven in Theorem~\ref{theorem1}. It indicates that the triangle area is a measure for genuine tripartite entanglement. This is further strengthened for qubits in Theorem~\ref{theorem3} that given a set of measures, the LOCC-monotonicity can be violated by three-qubit states for $\alpha>1/2$.}
	\label{Monotone}
\end{figure}

 \begin{Proposition}	\label{proposition2}
 	For any subadditive measure $\mathcal{E}$, the triangle area~(\ref{area}) enclosed by the relation~(\ref{triangle}) with $\alpha\in(0,1)$ is nonzero if and only if the pure tripartite state is genuinely entangled.
  \end{Proposition}
 
We prove Proposition~\ref{theorem2} by contradiction. Indeed, it equates to proving that the triangle relation~(\ref{triangle}) with $\alpha\in(0,1)$ can never be saturated by genuinely entangled states with three nonzero $\mathcal{E}^\alpha_{i|jk}$. If $\mathcal{E}^\alpha_{i|jk}=\mathcal{E}^\alpha_{j|ik}+\mathcal{E}^\alpha_{k|ij}$ holds for some $\alpha \in (0,1)$ and nonzero $\mathcal{E}^\alpha$, then
 \begin{align}
 	\mathcal{E}_{i|jk}&=(\mathcal{E}^\alpha_{i|jk})^{1/\alpha}=(\mathcal{E}^\alpha_{j|ik}+\mathcal{E}^\alpha_{k|ij})^{1/\alpha} \nn \\
 	&> (\mathcal{E}^\alpha_{j|ik})^{1/\alpha}+(\mathcal{E}^\alpha_{k|ij})^{1/\alpha}=	\mathcal{E}_{j|ik}+	\mathcal{E}_{k|ij}. \label{contradiction}
 	\end{align}
 The first inequality follows from $(x+y)^r> x^r+y^r$ for positive $x,y$ and $r=1/\alpha>1$. It is obvious that Eq.~(\ref{contradiction}) contradicts the triangle relation~\ref{triangle}. Thus, we complete the proof of Proposition~\ref{theorem2}, which provides a faithful geometric picture for genuinely entangled states. 
 
It is remarked that whether the triangle inequality~(\ref{triangle}) with $\alpha=1$ can be saturated depends on the state space and the measure. For example, there is $\mathcal{S}_{A|BC}=2$ and $\mathcal{S}_{B|AC}=\mathcal{S}_{C|AB}=1$  for genuinely entangled state $ (\ket{000}+\ket{101}+\ket{210}+\ket{311})/2$ and von Neumann entropy $\mathcal{S}$, while equality can never be achieved by genuinely entangled three-qubit states and Tsallis entropy~(\ref{Tsallis}) with $q>1$ in the Appendix~\ref{entanglementtraingle}. 

Proposition~\ref{theorem2} indicates that the triangle area~(\ref{area}) is a natural quantifier for pure tripartite entanglement. For a general state $\rho$, using the convex-roof construction
\beq\label{mixed}
\mathcal{A}(\rho):={\rm inf}_{\{p_m,\ket{\psi_m}\}}\sum_m p_m\,\mathcal{A}(\ket{\psi_m})
\eeq
where the infimum is over all pure decompositions $\rho=\sum_m p_m\ket{\psi_m}\bra{\psi_m}$, one can show that $\mathcal{A}(\rho)=0$ if and only if $\rho$ is biseparable, admitting a decomposition of which all pure states are biseparable. This implies the triangle area is a faithful quantifier of genuine tripartite entanglement for both pure and mixed states.

\section{Triangle area as an entanglement measure}\label{areameasure}

We continue to derive a stronger result that the triangle area is a measure for genuine tripartite entanglement.

\begin{Theorem}\label{theorem1}
	For any subadditive measure $\mathcal{E}$, the triangle area~(\ref{area}) with $\alpha\in(0,1/2]$ admits LOCC-monotonicity and hence is a reliable entanglement measure.
\end{Theorem}

Before proceeding to prove Theorem~\ref{theorem1}, we first introduce the parametrised vector $\mathbf{x}=(x_1, x_2, x_3)^\top=(\mathcal{E}^{2\alpha}_{i|jk},\mathcal{E}^{2\alpha}_{j|ik},\mathcal{E}^{2\alpha}_{k|ij})^\top$. Correspondingly, the area~(\ref{area}) can be rewritten as
\beq
f(\mathbf{x})=\frac{1}{4}\rt{-x^2_1+2x_1(x_2+x_3)-(x_2-x_3)^2}. \label{newform}
\eeq
Evidently, the function $f$ is continuous and permutation-invariant under parameters $x_i$. Moreover, we have

\begin{Lemma}\label{lemma1}
	For $\alpha\in[0,1/2]$, Eq.~(\ref{newform}) is nondecreasing and concave as a function of $(x_1, x_2, x_3)^\top=(\mathcal{E}^{2\alpha}_{i|jk},\mathcal{E}^{2\alpha}_{j|ik},\mathcal{E}^{2\alpha}_{k|ij})^\top$. 
\end{Lemma}

The nondecreasing tendency of $f$ over each $x_i$ is determined by its nonnegative first derivatives 
\beq\label{nonobtuse}
\frac{\partial f}{\partial x_i}=\frac{\partial \mathcal{A}}{\partial \mathcal{E}^{2\alpha}_{i|jk}}=\frac{\mathcal{E}^{2\alpha}_{j|ik}+\mathcal{E}^{2\alpha}_{k|ij}-\mathcal{E}^{2\alpha}_{i|jk}}{16\mathcal{A}} \geq 0.
\eeq
The inequality follows from Proposition~\ref{theorem1} that $x_1\leq x_2+x_3, x_2\leq x_1+x_3,$ and $x_3\leq x_1+x_2$ for $\alpha\leq 1/2$. It immediately yields that the triangle enclosed by~(\ref{triangle}) with $\alpha\in [0, 1/2]$ is non-obtuse, as its interior angles obey $\cos\theta_i=(\mathcal{E}^{2\alpha}_{j|ik}+\mathcal{E}^{2\alpha}_{k|ij}-\mathcal{E}^{2\alpha}_{i|jk})/2\mathcal{E}^{\alpha}_{j|ik}\mathcal{E}^{\alpha}_{k|ij}\geq 0$. The concavity of $f$ is determined by its Hessian matrix which is shown to be nonpositive definite in the Appendix~\ref{measures}. 

We then use Lemma~\ref{lemma1} to obtain the proof to LOCC-monotonicity of the triangle area, as displayed in Fig.~\ref{Monotone}. Being restricted to the pure state $\ket{\psi}$ and any pure LOCC-ensemble $\{p_m, \ket{\psi_m}\}$, we have
\begin{align}
	\sum_m p_m\, \mathcal{A}(\ket{\psi_m})& =\sum_m p_m\,f(\mathbf{x}^m) \leq  f(\sum_m p_m\,\mathbf{x}^m) \nn\\
	&= f(\sum_m p_m\,x_1^m, \sum_m p_m\,x_2^m, \sum_m p_m\,x_3^m)\nn \\
	&\leq f(x_1, x_2, x_3)=f(\mathbf{x})=\mathcal{A}(\ket{\psi}). \label{strongmonotonocity}
	\end{align}
All equalities follow directly from Eq.~(\ref{newform}) by associating each pure state with a parametrised vector $\mathbf{x}^m$, the first inequality from Lemma~\ref{lemma1}, and the second from Lemma~\ref{lemma1} and the fact that $\mathcal{E}^{2\alpha}_{i}$ is a measure of bipartite entanglement for $\alpha\leq 1/2$, i.e., $x_i\geq \sum_m p_m\,x^m_i$ for $i=1,2,3$. For a general state and its general LOCC ensemble, using the convex-roof rule~(\ref{mixed}) and thus convexity of the area leads to a similar proof of LOCC-monotonicity. Thus, we complete the proof of Theorem~\ref{theorem1}.

Note that as the proof of Theorem~\ref{theorem1} is independent of both state and measure, we obtain a unifying method to construct the faithful measure for genuine tripartite entanglement universally valid for discrete, continuous, and even hybrid quantum systems. In particular, we provide an example via the Tsallis entropy~(\ref{Tsallis}) with $q=2$ to confirm this.

\begin{Theorem}\label{theorem2.1}
	The triangle relation as per~(\ref{triangle}), in terms of Tsallis-2 entropy, is universal in sense that it is valid for all discrete, all Gaussian, and all discrete-discrete-continuous pure tripartite states. Moreover, the triangle area~(\ref{area}) with $\alpha\in(0,1/2]$ is a faithful measure for genuine tripartite entanglement.
	
\end{Theorem}

The length proof is deferred to the Appendix~\ref{sec3}. We remark that it is the first measure applicable for both general discrete systems and Gaussian systems, completely going beyond previous works which are restricted to either discrete systems or continuous ones, such as Refs.~\cite{Beckey2021,Guo2022} specialized for qubits and Ref.~\cite{Adesso2012} specialized for Gaussian states.  Note also that for $\alpha>1/2$, neither the nondecreasing nor concavity of $f$ can be always satisfied, signalling the possibility of violating LOCC-monotonicity and thus the triangle area may be not a measure. This is confirmed in the Appendix that with Tsallis entropy, the triangle area with $\alpha>1/2$ can be increased by LOCC on a family of W-class states.

\section{Upper and lower bounds}\label{bounds}

Following again from Lemma~\ref{lemma1} that the area is nondecreasing under each $\mathcal{E}^\alpha_{i|jk}$, we can obtain a lower bound
\beq \label{lower}
\mathcal{A}(\ket{\psi}_{ijk})\geq \min\frac{\sqrt{3}}{4}\{\mathcal{E}^{2\alpha}_{i|jk}, \mathcal{E}^{2\alpha}_{j|ik},\mathcal{E}^{2\alpha}_{k|ij}\},
\eeq
which can be interpreted as proportional to the squared smallest side of the triangle. This bound is also a measure for genuine tripartite entanglement. Especially, if $\mathcal{E}^\alpha$ is Tsallis-$2$ entropy with $\alpha=1/2$, then it recovers the well-known genuinely multipartite concurrence (GMC)~\cite{Ma2011}. Additionally, the triangle area is upper bounded by
\beq \label{upper}
\mathcal{A}(\ket{\psi}_{ijk})\leq\rt{Q\left(\frac{3Q-2Q}{3}\right)^3}\leq\frac{\mathcal{E}^{2\alpha}_{i|jk}+\mathcal{E}^{2\alpha}_{j|ik}+\mathcal{E}^{2\alpha}_{k|ij}}{4\sqrt{3}},
\eeq
proportional to the average of squared sides. The first inequality follows from $xyz\leq (\frac{x+y+z}{3})^3$ and the second from $(x+y+z)^2\leq 3(x^2+y^2+z^2)$ for nonnegative $x, y, z$. In terms of Tsallis-$2$ entropy, it reduces to the global entanglement measure for $\alpha=1/2$~\cite{Meyer2002,Brennen2003} which, however, can be nonzero even if the state is biseparable.

 \begin{table}[!hbp]
	\centering
	\begin{tabular}{c|c|c|c}
		\hline
		\hline
		& $\ket{\psi_1}$&  $\ket{\psi_2}$  & $\ket{\psi_3}$\\
		\hline
		GMC & $0.5878$& $0.7071$& $0.7071$\\
		$\mathcal{A}_1$ & $0.7329$& $0.6009$& $0.8251$  \\
		$\mathcal{A}_2$ &$0.6487$ & $0.5$ & $0.7638$\\
		\hline
		\hline
	\end{tabular}
	\caption{Genuine tripartite entanglement measured by triangle areas and GMC. $\mathcal{A}_1$ describes the triangle area of Tsallis entropy with $q=1$ and $\mathcal{A}_2$ the area of Tsallis entropy with $q=2$, with $\alpha=1/2$. The concurrence is used in GMC. A normalisation coefficient $(16/3)^{1/2}$ is applied to the triangle area to guarantee $\mathcal{A}_1,\mathcal{A}_2\leq1$ for all $3$-qubit states. } 
	\label{Tab1}
\end{table}

 We exemplify the main differences between the geometric measures and GMC for genuine tripartite entanglement. Denote $\mathcal{A}_1$ by the triangle area of Tsallis entropy with $q=1$, equivalent to von Neumann entropy, and $\mathcal{A}_2$ by the area of Tsallis entropy with $q=2$, with $\alpha=1/2$. The bipartite measure in GMC refers to concurrence. We in particular consider three states $ \ket{\psi_1}=(\sin\frac{\pi}{5}\ket{000}+\cos\frac{\pi}{5}\ket{100}+\ket{111})/\rt{2}, \ket{\psi_2}=\cos\frac{\pi}{8}\ket{000}+\sin\frac{\pi}{8}\ket{111}, \ket{\psi_3}=\frac{1}{2}\ket{000}+\frac{1}{2}\ket{100}+\frac{1}{\sqrt{2}}\ket{111}.$ It is shown in the rows of Table~\ref{Tab1} that two triangle areas have different entanglement orderings of three states, in comparison to GMC, while different columns of Table~\ref{Tab1} indicate that these three measures lead to different entanglement orderings of three states.

\section{Strengthened results for qubits}\label{qubits}

 When it is restricted to three-qubit states, all above results can be significantly strengthened. Here we consider the subadditive measures such as von Neumann entropy $\mathcal{S}$, Tsallis entropy~$\mathcal{T}$, squared concurrence $\mathcal{C}^2$, squared negativity $\mathcal{N}^2$, and non-subadditive ones, including Schmidt weight $\mathcal{W}$ and R\'enyi-$2$ entropy $\mathcal{R}$. It is shown in the Appendix~\ref{bipartitemeasures} that they can be unified via some function on the smallest eigenvalue $\lambda$ of the reduced state for any pure two-qubit state
 \begin{align}
 		\mathcal{S}(\lambda)&=-\lambda\log_2\lambda-(1-\lambda)\log_2(1-\lambda), \nn \\
 		\mathcal{T}(\lambda)&=\frac{1-\lambda^q-(1-\lambda)^q}{q-1},~ \mathcal{C}^2(\lambda)=\mathcal{N}^2(\lambda)=4\lambda(1-\lambda), \nn \\
   	\mathcal{W}(\lambda)&= 2\lambda,~\mathcal{R}(\lambda)=-\log\,[\lambda^2+(1-\lambda)^2].
\end{align}

Consequently, we can derive the following results, for which the proofs are deferred to the Appendices. 

\begin{Theorem}\label{theorem2}
	For the measure set  $\{\mathcal{S}, \mathcal{T}, \mathcal{C}^2, \mathcal{N}^2, \mathcal{W}, \mathcal{R}\}$, the triangle relation~(\ref{triangle}) holds for any $\alpha\in[0,1]$ on all pure three-qubit states, and can be violated by some state for any $\alpha>1$. Moreover, the triangle area~(\ref{area}) is nonzero if and only if the three-qubit state is genuinely entangled, except for Schmidt weight with $\alpha=1$.
\end{Theorem}

Theorem~\ref{theorem2} immediately yields that the non-subadditive measures, such as Schmidt weight and R\'enyi-$2$ entropy, are subadditive on all two-qubit states with rank no larger than $2$. It also strengthens Proposition~\ref{theorem2}, in the sense that $\alpha=1$ optimally upper bounds the triangle relation~(\ref{triangle}) for three-qubit states, and the faithful geometric picture is extended to the bound $\alpha=1$ for Tsallis entropy. Additionally, the triangle relation with $\alpha=1$ recovers the ones already obtained in~\cite{Zhu2015,Qian2018}, and reduces to the entanglement polytopes~\cite{Higuchi2003,Walter2013} in context of Schmidt weight. 

\begin{Theorem}\label{theorem3}
 	For the measure set $\{\mathcal{W}, \mathcal{C}^2, \mathcal{N}^2, \mathcal{S}, \mathcal{T},\mathcal{R}\}$, the triangle area~(\ref{area}) with $\alpha\in(0, 1/2]$ is an entanglement measure for three-qubit states, while it is not for $\alpha>1/2$.
\end{Theorem}

We note that violating the LOCC-monotonicity by the area induced by subadditive measures with $\alpha>1/2$ naturally implies the same violation for generic tripartite systems in Theorem~\ref{theorem1}. Moreover, Theorem~\ref{theorem3} rigorously confirms the concurrence fill~\cite{Ge2023,Jin2023} and the ergotropic fill~\cite{Puliyil2022} as feasible entanglement measures. It is also found in the Appendix that the proof in~\cite{Jin2023} is incomplete to guarantee the LOCC-monotonicity of the concurrence area.

\section{Discussion}\label{discussions}

We have presented an unified geometric picture suitable to characterise tripartite entanglement of discrete, continuous, and even hybrid quantum systems, and then proposed using the triangle area as a faithful measure for genuine tripartite entanglement. We have also obtained useful lower and upper bounds for these geometric measures, and explored their connections and differences with the well-known measures for multipartite entanglement. Especially, our results are significantly strengthened for qubits, which also generalise previous results and solve open questions left in previous works.  

Generalisations of our results are given as follows. Regard to LOCC-monotonicity, it follows from the convexity that the triangle area~(\ref{area}) with $\alpha\in (0, 1/2]$ also admits a weaker monotonicity in the form of $\mathcal{A}(\rho)\geq \sum_m\mathcal{A}(\sum_mp_m\rho_m)$. It is thus interesting to investigate whether our results can be applied to the measures only satisfying this weaker LOCC-monotonicity, i.e., $\mathcal{E}(\rho)\geq \sum_m\mathcal{E}(\sum_mp_m\rho_m)$. If the measure is non-faithful, i.e., $\mathcal{E}(\rho)=0 $ for some entangled state $\rho$, the corresponding triangle area can still be a measure for tripartite entanglement, but may not be faithful any more. For any non-subadditive measure $\mathcal{E}$, it has been shown in~\cite{Guo2022} that it always satisfies the triangle relation~(\ref{triangle}) for some $0<\beta<+\infty$, implying $\mathcal{E}^{\beta}$ is subadditive. It follows immediately from Lemma~\ref{lemma1} that $\mathcal{E}^\alpha$ with $\alpha\in (0, \beta/2]$ satisfies the non-obtuse condition~(\ref{nonobtuse}) and the enclosing area is a measure for genuine tripartite entanglement. However, it could be challenging to obtain a proper $\beta$ for the non-subadditive measure.

Finally, we point out that the triangle relation~(\ref{triangle}) can be generalized to a polygon relation
\beq
\mathcal{E}^\alpha_{A_1|\bar {A}_1} \leq \mathcal{E}^\alpha_{A_2|\bar {A}_2}+\dots \mathcal{E}^\alpha_{A_n|\bar {A}_n}
\eeq
 for both discrete and continuous $n$-partite states, with $A_i|\bar{A}_i$ denoting the bipartition $A_i$ and the rest parties. Hence, we expect our results to aid significant progress in studying entanglement of multipartite systems~\cite{Erhard2020}. We also hope these results find applications in studying other multipartite quantum resources, such as genuine nonlocality~\cite{Tavakoli2022} and steering~\cite{Xiang2022}.

\acknowledgements~We greatly thank Dr. Michael Hall for fruitful discussions and useful suggestions. This work is supported by the Shanghai Municipal Science and Technology Fundamental Project (No. 21JC1405400), the Fundamental Research Funds for the Central Universities (No. 22120230035), the National Natural Science Foundation of China (No. 12205219,62173288), the Shanghai Municipal Science and Technology Major Project (2021SHZDZX0100), the Hong Kong Research Grant Council (No. 15203619,15208418), and the Shenzhen Fundamental Research Fund (No. JCYJ20190813165207290).\\


\appendix

\section{Entanglement measures for the bipartite system}\label{bipartitemeasures}

Any pure two-qubit state $\ket{\psi}_{AB}$, shared by two parties $A$ and $B$, admits the Schmidt decomposition as~\cite{Nielsen2000}
\beq\label{twoqubitstate}
\ket{\psi}_{AB}=\sqrt{\lambda_1}\,\ket{00}+\sqrt{\lambda_2}\,\ket{11},
\eeq
where the Schmidt coefficients satisfies $\lambda_1+\lambda_2=1$, and $\ket{0}, \ket{1}$ denote the computational basis for each subsystem. It is noted that the state (\ref{twoqubitstate}) is entangled if and only if $\lambda_1\lambda_2\neq0$, and the degree of entanglement can be quantified via entanglement measures, which are typically defined as a certain function $\mathcal{E}$ mapping a general state to a real number in the interval $[0,1]$. Indeed, one possible measure needs to satisfy some basic constraints, like being zero for non-entangled states and being invariant under local unitary operations. Further, it is natural to be non-increasing under LOCC, as these operations can be freely implemented within the resource theory of entanglement. Numerous measures conforming the above requirements have been proposed to quantify bipartite entanglement.

We then exemplify several well-known entanglement measures to be used in the subsequent sections. Denote the reduced state $\rho_i={\rm Tr}_{j}[\ket{\psi}_{AB}\bra{\psi}]$ for $i,j=A,B$, and it is evident that $\lambda_1$ and $\lambda_2$ correspond to two eigenvalues of both $\rho_A$ and $\rho_B$. Since the Schmidt coefficients fully determine whether the state~(\ref{twoqubitstate}) is entangled or not, it is straightforward to introduce the Schmidt weight~\cite{Grobe1994}
\beq\label{schmidtweight}
\mathcal{W}(\ket{\psi}_{AB})= 2\min\{\lambda_1,\lambda_2 \}
\eeq
as one entanglement measure. It is remarked that this measure also has an operational interpretation as the ergotropic gap in quantum thermodynamics~\cite{Puliyil2022}. The second one is the concurrence~\cite{Wootters1998,Rungta2001,Albeverio2001}
\beq\label{concurrence}
\mathcal{C}(\ket{\psi}_{AB})=\sqrt{2(1-{\rm Tr}[\rho_i^2])}=2\sqrt{\lambda_1\lambda_2}.
\eeq
For pure states, it is completely equivalent to another commonly-used measure, called the negativity $\mathcal{N}$~\cite{Vidal2002}. It follows further from the first equality in~Eq.~(\ref{concurrence}) that the concurrence can be regarded as a special class of square-root quantum Tsallis entropy $(1-\tr{\rho^q})/(1-q)$ for the reduced states~\cite{Tsallis1988,Hu2006}. The other information-theoretical measures for entanglement are the von Neumann entropy of the reduced state~\cite{Nielsen2000}
\begin{align}\label{vonentropy}
	\mathcal{S}(\ket{\psi}_{AB})=-{\rm Tr}[\rho_i\log\rho_i]=-\lambda_1\log\lambda_1-\lambda_2\log\lambda_2,
\end{align}
and the R\'enyi-$2$ entropy
\begin{align}\label{Renyientropy}
	\mathcal{R}(\ket{\psi}_{AB})=-\log_2{\rm Tr}[\rho^2]=-\log_2(\lambda_1^2+\lambda_2^2).
\end{align}

Finally, it is interesting to observe that entanglement measures for the pure two-qubit state~(\ref{twoqubitstate}) can be unified as the function $\mathcal{E}(\lambda)$ that is strictly increasing and concave over the smallest eigenvalue of the reduced state $\lambda=\min\{\lambda_1,\lambda_2 \}$. For examples, the above mentioned entanglement measures are equivalent to
\begin{eqnarray}
	\mathcal{W}(\lambda)&= &2\lambda, \label{weightlambda} \\
	\mathcal{C}^2(\lambda)&=&4\lambda(1-\lambda),\label{concurrencelambda} \\
	\mathcal{N}^2(\lambda)&=&4\lambda(1-\lambda),\label{negativitylambda} \\
	\mathcal{S}(\lambda)&=&-\lambda\log_2\lambda-(1-\lambda)\log_2(1-\lambda),\label{von}\\
	\mathcal{T}(\lambda)&=&\frac{1-\lambda^q-(1-\lambda)^q}{q-1}.\label{tallis}\\
	\mathcal{R}(\lambda)&=&-\log_2(\lambda^2+(1-\lambda)^2).\label{renyilambda}
\end{eqnarray}
Thus, it is relatively easy to quantify entanglement for the two-qubit system and even general bipartite systems. However, multipartite states generically do not admit the Schmidt decomposition as simple as Eq.~(\ref{twoqubitstate}). Besides, the state structure allows for the partial entanglement and genuine entanglement and the mathematical structure of multipartite LOCC is rather complex, thus making it challenging to quantify multipartite entanglement via proper measures.

\section{Bipartition of the multipartite system}\label{multisystem}

In order to faithfully distinguish partial entanglement and genuine multipartite entanglement, one potential measure needs faithfulness that if a multipartite state is biseparable, i.e., there exists at least one bipartition of the state being separable, then it is zero; Otherwise, it is strictly positive. This immediately indicates that whether the state is genuinely entangled depends on bipartite entanglement generated from the multipartite state.
Here, we are restricted to the tripartite system and illustrate the idea about using the well-explored bipartite entanglement measures to study the nonclassical property of multipartite quantum states.  

Particularly, an arbitrary tripartite state, shared by three parties $A, B,$ and $C$, has three bipartition: $A|BC$,  $B|AC$, and $C|AB$. Consequently, these bipartite states can be quantified by the entanglement measures introduced in the above subsection, and then we are left to check whether three $\mathcal{E}_{i|jk}$, measuring the entanglement between bipartition $i$ and $jk$ with $i,j,k=A,B,C$, is zero or not.

For examples, noting that the three-qubit generalised Greenberger-Horne-Zeilinger (GHZ) states
\beq
\ket{{\rm GHZ}}=\cos\theta\,\ket{000}+\sin\theta\,\ket{111}, ~~ \theta\in [0, \pi/4], \label{GHZ}
\eeq
are symmetric under the party permutations, $\mathcal{E}_{i|jk}$ are identical to each other and is nonzero for $\theta\neq 0$, implying they are genuinely entangled. Another class of genuinely entangled states is the generalised W state
\beq
\ket{{\rm W}}=a\ket{100}+b\ket{010}+c\ket{001},  \label{W}
\eeq
with nonzero coefficients satisfying $|a|^2+|b|^2+|c|^2=1$. It has been shown in~\cite{Dur2000} that they are two distinct classes of entangled three-qubit states, under the stochastic LOCC.

We has simply explained that the bipartite measures are useful for investigating the problem of whether a tripartite state is genuinely entangled or not. In the following sections, we go further to explore their utility in characterising and quantifying tripartite entanglement.

\section{Unifying  triangle relations for tripartite entanglement}\label{entanglementtraingle}

Given a pure three-qubit state $\ket{\psi}_{ABC}$, denote $\lambda^i=\min\{\lambda^i_1,\lambda^i_2 \}\in [0, 1/2]$ as the smallest eigenvalue of $\rho_i={\rm Tr}_{jk}[|\psi\rangle_{ijk}\langle\psi|]$, so we can introduce the measure $\mathcal{E}_{i|jk}(\lambda^i)$ to quantify its bipartite entanglement. Following from the purity of the state and the subadditivity of the entropies~\cite{Araki1970,Auden2007}, and the concavity of the functions Eq.~(\ref{weightlambda})-(\ref{negativitylambda}) for the Schmidt weight, squared concurrence, and squared negativity~\cite{Qian2018}, these bipartite entanglement measures satisfy the triangle inequality
\beq
\mathcal{E}_{i|jk} \leq \mathcal{E}_{j|ik}+\mathcal{E}_{k|ij}, \label{polygon}
\eeq
where $\lambda^i$ is dropped for simplicity.

Further, the above  relation can be generalised to
\beq
\mathcal{E}_{i|jk}^\alpha \leq \mathcal{E}_{j|ik}^\alpha+\mathcal{E}_{k|ij}^\alpha, ~~ \forall \alpha\in(0,1]. \label{triangle1}
\eeq
We provide a simple algebraic proof of the triangle relation (\ref{triangle1}), by showing 
\begin{eqnarray}
	\mathcal{E}^\alpha_{i|jk} 
	&\leq & (\mathcal{E}_{j|ik}+ \mathcal{E}_{k|ij})^{\alpha}\nn \\
	&\leq& \mathcal{E}^\alpha_{j|ik}+\mathcal{E}^\alpha_{k|ij},\label{triangleproof}
\end{eqnarray}
where the first inequality follows directly from Eq.~(\ref{polygon}), and the second is due to the relation $(a+b)^r\leq a^r+b^r$ for nonnegative $a, b$  and $r\leq 1$. Hence, we can conclude that the triangle relation~(\ref{triangle1}) is valid for $\alpha \in (0, 1]$. This also leads to the following useful corollary.
\begin{Corollary}
	If there is a positive $\alpha$ such that $\mathcal{E}^\alpha_{i|jk}\leq\mathcal{E}^\alpha_{j|ik}+\mathcal{E}^\alpha_{k|ij}$ for all pure states, then $\mathcal{E}^\beta_{i|jk}\leq\mathcal{E}^\beta_{j|ik}+\mathcal{E}^\beta_{k|ij}$ for any $0<\beta\leq\alpha$.
\end{Corollary}

Finally, to show that $\alpha=1$ is the optimal upper bound for the triangle relation~(\ref{triangle1}), we present some lemmas:

\begin{Lemma}\label{LemmaS1}
	Suppose the differentiable, and non-negative function $\mathcal{E}(\lambda)$ is strictly increasing and concave, i.e., $\mathcal{E}^{\prime}(\lambda)>0$ and $\mathcal{E}^{\prime\prime}(\lambda)\leq 0$. The function $\mathcal{E}^\alpha(\lambda)$ is strictly convex if and only if 
	\beq
	(\alpha-1)\mathcal{E}^{\prime2}(\lambda)+\mathcal{E}(\lambda)\mathcal{E}^{\prime\prime}(\lambda)> 0. \label{convexity}
	\eeq
\end{Lemma}

This Lemma is easy to be confirmed by requiring $ (\mathcal{E}^\alpha(\lambda))^{\prime\prime}> 0$ for the strict-convex property of $\mathcal{E}^\alpha(\lambda)$. It is also evident from Lemma~\ref{LemmaS1} that it is impossible for $\mathcal{E}^\alpha(\lambda)$ with $\alpha \leq 1$ to be strictly convex and thus likely to be convex only if $\alpha>1$. Then, there is

\begin{Lemma}\label{LemmaS2}
	Suppose the function $\mathcal{E}^\alpha(\lambda)$ is strictly convex in some non-empty interval $\Delta_{\alpha}=[0,u_{\alpha}]$ with $\mathcal{E}(0)=0$. If  $\lambda_1, \lambda_2$, and $\lambda_1+\lambda_2$ are in $\Delta_{\alpha}$, then we have
	\beq
	\mathcal{E}^{\alpha}(\lambda_1+\lambda_2)> 	\mathcal{E}^{\alpha}(\lambda_1)+\mathcal{E}^{\alpha}(\lambda_2). \label{counterexampleS1}
	\eeq
\end{Lemma}

Lemma~\ref{LemmaS2} follows directly from the strict convexity of $\mathcal{E}^\alpha(\lambda)$ that  
\begin{eqnarray*}
	\mathcal{E}^\alpha(\lambda_1+\lambda_2)
	&=& \frac{\lambda_1}{\lambda_1+\lambda_2}\mathcal{E}^\alpha(\lambda_1+\lambda_2)+\frac{\lambda_2}{\lambda_1+\lambda_2}\mathcal{E}^\alpha(0)\\
	&&+ \frac{\lambda_2}{\lambda_1+\lambda_2}\mathcal{E}^\alpha(\lambda_1+\lambda_2)+\frac{\lambda_1}{\lambda_1+\lambda_2}\mathcal{E}^\alpha(0)  \\
	&>& 	\mathcal{E}^\alpha(\lambda_1)+	\mathcal{E}^\alpha(\lambda_2).
\end{eqnarray*}  
In turn, it indicates that if $\mathcal{E}^\alpha$ is strictly convex, then it is possible to find states to satisfy Eq.~(\ref{counterexampleS1}) and hence to violate the triangle relation~(\ref{triangle1}). Thus, what is left to do is to check whether there exist physical states and the corresponding measures $\mathcal{E}$ such that $\mathcal{E}^\alpha$ is strictly convex.

Recall from Eq.~(\ref{weightlambda})-(\ref{tallis}) that these measures $\mathcal{E}$ can be unified as functions of $\lambda$. Specifically, for the Schmidt weight (\ref{weightlambda}), we have
\beq
(\mathcal{W}^\alpha(\lambda))^{\prime\prime}=(2^\alpha\lambda^\alpha)^{\prime\prime}=2^\alpha\alpha(\alpha-1)\lambda^{\alpha-2}. 
\eeq
It is evident that the second derivative of $\mathcal{W}^\alpha$ is always strictly positive for any $\alpha>1$, implying the strict convexity. Further, consider the W-class state~(\ref{W}) with $|a|^2\geq1/2$. It is easy to obtain 
\beq \label{threew}
\lambda^A=|b|^2+|c|^2, \lambda^B=|b|^2, \lambda^C=|c|^2. 
\eeq
Immediately, following from Lemma~\ref{LemmaS2}, one has
\beq
\mathcal{W}_{A|BC}^\alpha=(\mathcal{W}_{B|AC}+\mathcal{W}_{C|AB})^\alpha > \mathcal{W}_{B|AC}^\alpha+\mathcal{W}_{C|AB}^\alpha,	 
\eeq
for any $\alpha>1$. This indicates that $\alpha=1$ is the optimal upper bound for the triangle relation~(\ref{triangle1}) based on the Schmidt weight.

Similarly, for the squared concurrence~(\ref{concurrencelambda}), there is
\beq
[\mathcal{C}^{2\alpha}(\lambda)]^{\prime\prime}=4^\alpha\alpha[\lambda(1-\lambda)]^{\alpha-2}[(\alpha-1)(1-2\lambda)^2-2\lambda(1-\lambda)].
\eeq
It is positive in the interval $\Delta_{\alpha}=[0,\frac{1}{2}(1-\sqrt{\frac{1}{2\alpha-1}})]$, which is not empty for any $\alpha>1$. Again, using Eq.~(\ref{threew}) for the W-class state enable us to find the violation of the triangle relation for any $\alpha>1$.

\begin{figure*}
	\begin{center}
		\includegraphics[width= 1.0\textwidth]{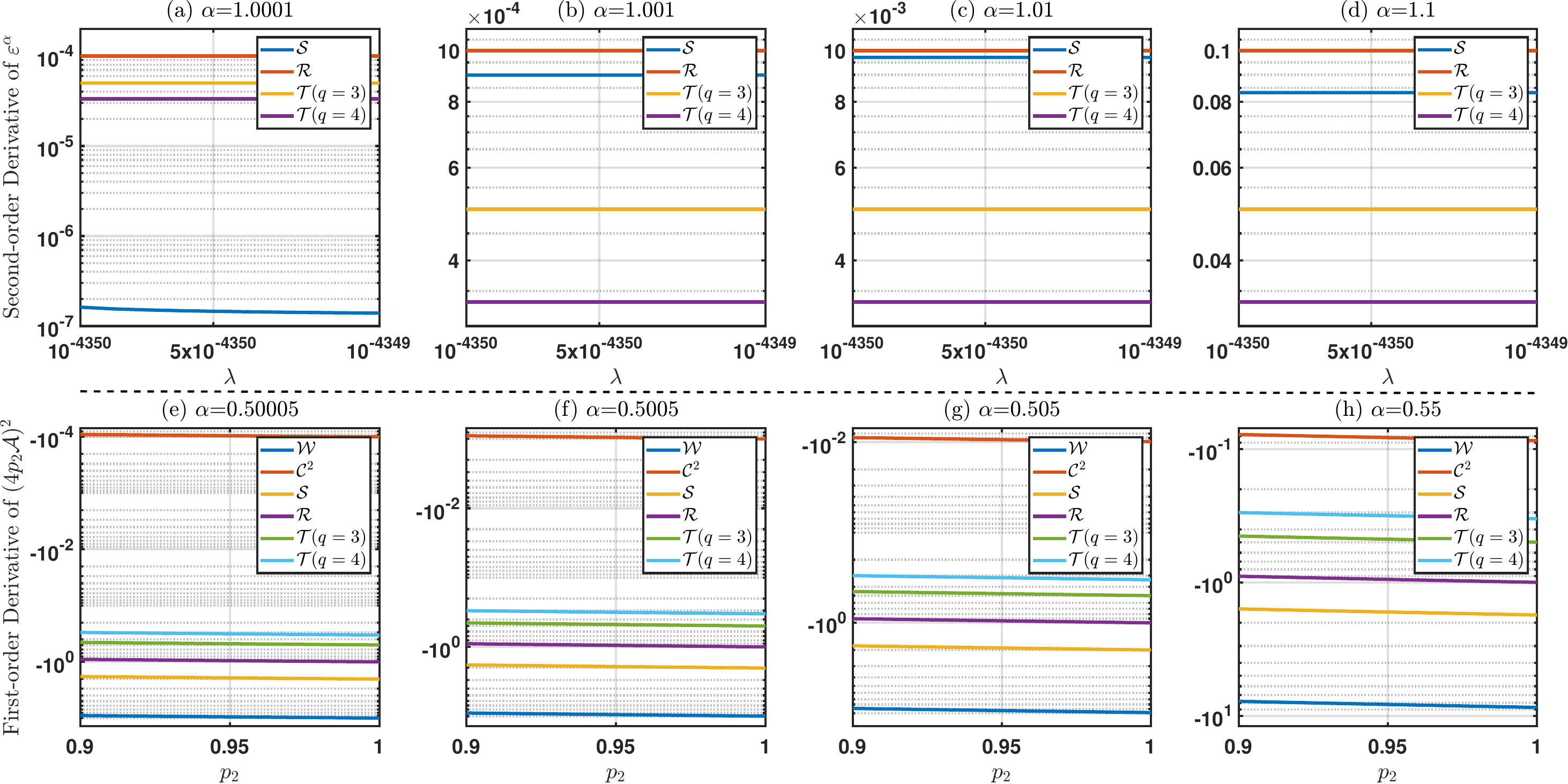}
	\end{center}
	\caption{Numerical results. In the upper, we plot $l_1(\lambda)/(2\alpha-1)$, $l_2(\lambda)/(q-1)$, and $l_3(\lambda)$ with different $\alpha$. Here, $l_i$ is a function that has the same sign with the second-order derivative of $\mathcal{E}^\alpha$ as shown in Eqs.~(\ref{vonl})-(\ref{Rl}). The results show that the second-order derivative of $\mathcal{E}^{\alpha}(\lambda)$ is positive in a interval when $\alpha>1$, implying the convexity of $\mathcal{E}^\alpha$. In the lower, with $\beta=10^{-30}$ and different $\alpha$ ($\alpha>1/2$), we plot $H(p_2)=L(\beta,p_2)\cdot k(\beta,p_2)$ with $k(\beta,p_2)>0$, which has the same sign with the first-order derivative of $g^2(p_2)$. Specifically, we set $K(\beta,p_2)=1$, $1/(2\alpha-1)$, $2/(2\alpha-1)$, $\mathcal{E}^{2\alpha+1}(2\beta/p_2)$, and $ \mathcal{E}^{2\alpha+1}(2\beta/p_2)/(q-1)$ for $\mathcal{C}^2$, $\mathcal{W}$, $\mathcal{S}$, $\mathcal{R}$, and $\mathcal{T}$, respectively. The results show that the first-order derivative of $g^2(p_2)$ are negative in the interval $[0.9,1]$, which implies that $g(p_2)$ is strictly decreasing.  }
	\label{numerical}
\end{figure*}

Regard to the entropic measures~(\ref{von})-(\ref{tallis}), we obtain 
\begin{eqnarray}
	\mathcal{S}^\alpha(\lambda)''&=&h_1(\lambda)\{(\alpha-1)[\log_2(1-\lambda)-\log_2\lambda]^2+(\ln2)^{-1}\nn\\
	&&[\log_2\lambda/(1-\lambda)+\log_2(1-\lambda)/\lambda] \}/(\log_2\lambda)^2\nn\\
	&\equiv& h_1(\lambda)l_1(\lambda)\label{vonl}
\end{eqnarray}
with $h_1(\lambda)=\alpha(\log_2\lambda)^2[-\lambda\log_2\lambda-(1-\lambda)\log_2(1-\lambda)]^{\alpha-2}$,
\begin{eqnarray}
	\mathcal{T}^\alpha(\lambda)''
	&=&h_2(\lambda)\big\{ (\alpha-1)[-\lambda^{q-1}
	+(1-\lambda)^{q-1}]^2-\nn\\
	&&(q-1)/q[1-\lambda^q-(1-\lambda)^q][\lambda^{q-2}+(1-\lambda^{q-2})] \big\}\nn\\
	&\equiv& h_2(\lambda)l_2(\lambda)\label{Tl}
\end{eqnarray}
with $h_2(\lambda)=\alpha q^2[1-\lambda^q-(1-\lambda)^q]^{\alpha-2}/(q-1)^\alpha$, and 
\begin{eqnarray}
	\mathcal{R}^\alpha(\lambda)''&=&h_3(\lambda)[(\alpha-1)(1-2\lambda)^2+2\ln2\cdot\lambda\nn\\
	&&(1-\lambda)\log_2(2\lambda^2-2\lambda+1)]\nn\\
	&\equiv& h_3(\lambda)l_3(\lambda)\label{Rl}
\end{eqnarray}
with $h_3=4\alpha(\ln_2)^{-2}[-\log_2(2\lambda^2-2\lambda+1)]^{\alpha-2}/(2\lambda^2-2\lambda+1)^2$. Noting $h_i(\lambda)>0$ for any $\lambda\in(0,1/2)$, and
\beq
\lim_{\lambda\rightarrow 0}l_i(\lambda)=\alpha-1>0, ~~i=1, 2, 3.
\eeq
Indeed, there must exist an interval $\Delta_\alpha=(0,u_\alpha]$ such that $(\mathcal{E}^\alpha(\lambda))''>0$ for any $\alpha>1$ (See Fig.~\ref{numerical} (a)-(d)), implying that $\mathcal{E}^\alpha(\lambda)$ is strictly convex in the interval $\Delta_\alpha$. Hence, $\alpha=1$ is also the optimal upper bound for the entropic triangle relations. 

\section{Faithful geometric picture for genuine tripartite entanglement}

It has been detailed in the main text that the triangle relation~(\ref{triangle1}) provides a nicely geometrical description of $\ket{\psi}_{ABC}$ that bipartite entanglement quantified by $\mathcal{E}^\alpha_{i|jk}$ can be interpreted as the side length of a triangle. When the state is biseparable, i.e., at one least zero $\mathcal{E}_{i|jk}$, then the triangle will degenerate to a line or a point (the product state case). By contrast, since all $\mathcal{E}_{i|jk}^\alpha\equiv\mathcal{E}^\alpha_{i|jk}(\lambda_i)\neq 0$ for genuinely entangled states, Theorem~5 of the main text is equivalent to that $\mathcal{E}^{\alpha}_{i|jk}$ must satisfy the strict triangle relation $\mathcal{E}_{i|jk}^\alpha< \mathcal{E}^\alpha_{j|ik}+\mathcal{E}^\alpha_{k|ij}$, or equivalently, the equality in~(\ref{triangle1}) is only achieved with at least one zero $\mathcal{E}^{\alpha}_{i|jk}$. 

Noting from the arguments in above subsection that the functions Eqs.~(\ref{weightlambda})-(\ref{tallis} are strictly concave, we can derive that the equality in the relation~(\ref{triangle1}) cannot be achieved if $\alpha\in(0,1)$, indicating the triangle relation~(\ref{triangle1}) with $\alpha\in(0,1)$ encloses a nondegenerate triangle for an arbitrary genuinely entangled state. However, this excludes the special case of the Schmidt weight with $\alpha=1$ with the evidence given by~(\ref{threew}).

We remark that the triangle, enclosed by the triangle relation with $\alpha \in (0, 1/2]$, is always non-obtuse, as each interior angle admits
\beq
\cos\theta_i=\frac{\mathcal{E}^{2\alpha}_{j|ik}+\mathcal{E}^{2\alpha}_{k|ij}-\mathcal{E}^{2\alpha}_{i|jk}}{2\mathcal{E}^{\alpha}_{j|ik}\mathcal{E}^{\alpha}_{k|ij}}\geq 0, \label{acute}
\eeq
for $2\alpha\in (0, 1]$. This condition plays a critical role in determining various useful properties of the triangle area and even whether it corresponds to a feasible genuine measure for tripartite entanglement.

\section{Unifying the triangle area as a measure for genuine tripartite entanglement}\label{measures}

Following the description above, it is natural to consider the triangle area, induced by the triangle relation~(\ref{triangle1}), 
\beq \label{area1}
\mathcal{A}(|\psi\rangle_{ijk})=\sqrt{Q(Q-\mathcal{E}^\alpha_{i|jk})(Q-\mathcal{E}^\alpha_{j|ik})(Q-\mathcal{E}^\alpha_{k|ij})}
\eeq
with the semiperimeter $Q=(\mathcal{E}^\alpha_{i|jk}+\mathcal{E}^\alpha_{j|ik}+\mathcal{E}^\alpha_{k|ij})/2$, as a promising entanglement measure for tripartite states. It can be normalised by multiplying $4/\rt{3}$, which is dropped for simplicity. It is easy to verify that $\mathcal{A}$ is invariant under local unitary operations, and is zero if and only if the state is biseparable for any $0<\alpha<1$. To determine whether it is a feasible entanglement measure or not, we need to examine its monotonicity under LOCC.

First, the area~(\ref{area1}) can be rewritten as
\beq \label{area2}
\mathcal{A}
=\frac{\rt{2\mathcal{E}^{2\alpha}_{i|jk}(\mathcal{E}^{2\alpha}_{j|ik}+\mathcal{E}^{2\alpha}_{k|ij})-\mathcal{E}^{4\alpha}_{i|jk}-(\mathcal{E}^{2\alpha}_{j|ik}-\mathcal{E}^{2\alpha}_{k|ij})^2}}{4}.
\eeq
Replacing the parameter vector $(\mathcal{E}^{2\alpha}_{i|jk},\mathcal{E}^{2\alpha}_{j|ik},\mathcal{E}^{2\alpha}_{k|ij})^\top$ with $(x_1, x_2, x_3)^\top$ immediately recovers Eq.~(10) of the main text. Further, we can obtain
\beq\label{gradient}
\frac{\partial \mathcal{A}}{\partial \mathcal{E}^{2\alpha}_{i|jk}}=\frac{\mathcal{E}_{j|ik}^{2\alpha}+\mathcal{E}_{k|ij}^{2\alpha}-\mathcal{E}_{i|jk}^{2\alpha}}{16\mathcal{A}}.
\eeq
It follows from the triangle relation~(\ref{triangle1}) that $\partial \mathcal{A}/\partial \mathcal{E}^{2\alpha}_{i|jk}\geq 0$ for any $0<\alpha\leq 1/2$, implying the area $\mathcal{A}$ is increasing under each $\mathcal{E}^{2\alpha}_{i|jk}$, while it is not always the case for $\alpha>1/2$. In the following, we discuss the LOCC-monotonicity of the triangle area in the cases $0<\alpha\leq1/2$, $1/2<\alpha<1$, and $\alpha=1$, respectively.

\subsubsection{Case I: $0<\alpha\leq 1/2$}
In the case $0<\alpha\leq1/2$, each $\mathcal{E}_{i|jk}^{2\alpha}$ is a measure for bipartite entanglement satisfying $\mathcal{E}_{i|jk}^{2\alpha}\geq\sum_m p_m \mathcal{E}_{i|jk}^{2\alpha(m)}$, where $\mathcal{E}^{2\alpha(m)}_{i|jk}$ denotes the corresponding bipartite entanglement of the state from any LOCC-ensemble $\{p_m,\rho_m\}$. However, it is possible that $\mathcal{E}^{2\alpha}_{i|jk}\leq \mathcal{E}_{i|jk}^{2\alpha(m)}$ and $\mathcal{A}(\mathcal{E}_{i|jk}^{2\alpha},\mathcal{E}_{j|ik}^{2\alpha},\mathcal{E}_{k|ij}^{2\alpha})\leq \mathcal{A}(\mathcal{E}_{i|jk}^{2\alpha(m)},\mathcal{E}_{j|ik}^{2\alpha(m)},\mathcal{E}_{k|ij}^{2\alpha(m)})$ for certain $m$. For example, consider the W-class state $\ket{{\rm W}}=\frac{\sqrt{3}}{2}\ket{100}+\frac{\sqrt{2}}{4}(\ket{010}+\ket{001})$, and that the local measurement $\{X_1,X_2 \}$ is acted on the first qubit with $X_1=\begin{bmatrix}
	\sqrt{3}/2 &0\\
	0 & \sqrt{2}/2
\end{bmatrix}$ and $X_2=\begin{bmatrix}
	1/2 &0\\
	0 & \sqrt{2}/2
\end{bmatrix}$. Then we have $\ket{{\rm W}^{(1)}}=\frac{\sqrt{6}}{3}\ket{100}+\frac{\sqrt{6}}{6}(\ket{010}+\ket{001})$. Since
\beq
\lambda^A=\frac{1}{4},~~\lambda^B=\lambda^C=\frac{1}{8}
\eeq
for the state $\ket{{\rm W}}$, and 
\beq
\lambda^{A}=\frac{1}{3},~~\lambda^B=\lambda^C=\frac{1}{6}
\eeq
for $\ket{{\rm W}^{(1)}}$, there is $\mathcal{A}(\ket{\rm W})<\mathcal{A}(\ket{{\rm W}^{(1)}})$. This indicates that the monotonicity of the triangle area only cannot guarantee its LOCC-monotonicity, and the proof in \cite{Jin2023} is incomplete.

However, if $\mathcal{A}$ is further concave under $(\mathcal{E}^{2\alpha}_{i|jk},\mathcal{E}^{2\alpha}_{j|ik},\mathcal{E}^{2\alpha}_{k|ij})^\top$, then the area $\mathcal{A}$ obeys the strong LOCC-monotonicity as
\begin{eqnarray}
	&&\mathcal{A}(\mathcal{E}^{2\alpha}_{i|jk},\mathcal{E}^{2\alpha}_{j|ik},\mathcal{E}^{2\alpha}_{k|ij})\nn \\
	&\geq&\mathcal{A}(\sum_mp_m\mathcal{E}^{2\alpha(m)}_{i|jk}, \sum_mp_m\mathcal{E}^{2\alpha(m)}_{j|ik}, \sum_mp_m\mathcal{E}^{2\alpha(m)}_{k|ij})\nn \\
	&\geq& \sum_m p_m \mathcal{A}(\mathcal{E}^{2\alpha(m)}_{i|jk}, \mathcal{E}^{2\alpha(m)}_{j|ik}, \mathcal{E}^{2\alpha(m)}_{k|ij}), \label{monotonicity2}
\end{eqnarray}
where $(\mathcal{E}^{2\alpha(m)}_{i|jk}, \mathcal{E}^{2\alpha(m)}_{j|ik}, \mathcal{E}^{2\alpha(m)}_{k|ij} )^\top$ corresponds to the parameter vector of each state from any LOCC-ensemble $\{p_m, \rho_m\}$. The first inequality follows from the monotonicity of $\mathcal{A}$, and the second inequality follows from the concavity of $\mathcal{A}$. Next, we prove Lemma~1 of the main text which ensures the validity of the above equation~(\ref{monotonicity2}). 

By tedious calculation, we can obtain that the Hessian matrix of $\mathcal{A}$ under the vector $(x_1,x_2,x_3)^\top=(\mathcal{E}^{2\alpha}_{A|BC}, \mathcal{E}^{2\alpha}_{B|AC},\mathcal{E}^{2\alpha}_{C|AB})^\top$ is 
\beq\label{Hessian}
\begin{bmatrix}
	-2x_2x_3 & x_3(x_1+x_2-x_3) & x_2(x_1+x_3-x_2)\\
	x_3(x_1+x_2-x_3) & -2x_1x_3 & x_1(x_2+x_3-x_1)\\
	x_2(x_1+x_3-x_2) & x_1(x_2+x_3-x_1) & -2x_1x_2
\end{bmatrix},
\eeq
where a positive coefficient $1/128\mathcal{A}^3$ is omitted for simplicity. Denote its $s$th order sequential principal minor by $D_s$ with $s=1,2$, and $3$. Then there is 
\begin{eqnarray*}
	D_1&=&-2x_2x_3\leq0,\\
	D_2&=&\det\left(\begin{bmatrix}
		-2x_2x_3 & x_3(x_1+x_2-x_3)\\
		x_3(x_1+x_2-x_3) & -2x_1x_3
	\end{bmatrix}\right)\\
	&=&x_3^2[x_1(x_2+x_3-x_1)+x_2(x_1+x_3-x_2)+x_3\cdot\\
	&&(x_1+x_2-x_3)]\geq 0,\\
	D_3&=& \det(H)=0,
\end{eqnarray*}
which implies that the Hessian matrix is negative semidefinite. Hence, we can obtain the area function $\mathcal{A}$ is concave under the parameter vector. This further enables us to obtain the first part of Theorem~3 and Theorem~6 as proven in the main text.

It is worth noting that if the vector $\mathbf{x}$ is instead denoted by $(\mathcal{E}^{\alpha}_{i|jk},\mathcal{E}^{\alpha}_{j|ik},\mathcal{E}^{\alpha}_{k|ij})^\top$, then the corresponding Hessian matrix $H$ is given by
\bqa
H_{11}&=&\frac{-x_1^6-3x_1^2(x_2^2-x_3^2)^2+(x_2^2+x_3^2)[3x_1^4+(x_2^2-x_3^2)^2]}{128\mathcal{A}^3},\nn \\
H_{22}&=&\frac{-x_2^6-3x_2^2(x_1^2-x_3^2)^2+(x_1^2+x_3^2)[3x_2^4+(x_1^2-x_3^2)^2]}{128\mathcal{A}^3},\nn \\
H_{33}&=&\frac{-x_3^6-3x_3^2(x_1^2-x_2^2)^2+(x_1^2+x_2^2)[3x_3^4+(x_1^2-x_2^2)^2]}{128\mathcal{A}^3},\nn \\
H_{12}&=& H_{21}= x_1x_2x_3^2(x_3^2-x_1^2-x_2^2)/(32\mathcal{A}^3),\nn\\
H_{13}&=&H_{31}= x_1x_2^2x_3(x_2^2-x_1^2-x_3^2)/(32\mathcal{A}^3),\nn\\
H_{23}&=&H_{32}= x_1^2x_2x_3(x_1^2-x_2^2-x_3^2)/(32\mathcal{A}^3).\nn
\eqa
The Hessian matrix is not negative semidefinite because 
\beq
\det(H)=\frac{x_1^2+x_2^2+x_3^2}{32\mathcal{A}}\geq0.
\eeq
Thus, it is impossible to obtain the concavity of the area.

\newcounter{mytempeqncnt}
\begin{figure*}[!t]
	\normalsize
	\setcounter{mytempeqncnt}{\value{equation}}
	\setcounter{equation}{29}
	\begin{equation}\label{Hessian2}
		\frac{1}{128\mathcal{A}^3}\left[
		\begin{array}{ccc}
			-x_1^6-3x_1^2(x_2^2-x_3^2)^2+[3x_1^4+(x_2^2-x_3^2)^2](x_2^2+x_3^2) & 4x_1x_2x_3^2(x_3^2-x_1^2-x_2^2)& 4x_1x_2^2x_3(x_2^2-x_1^2-x_3^2)\\
			
			4x_1x_2x_3^2(x_3^2-x_1^2-x_2^2) &  -x_2^6-3x_2^2(x_1^2-x_3^2)^2+[3x_2^4+(x_1^2-x_3^2)^2](x_1^2+x_3^2)   & 4x_1^2x_2x_3(x_1^2-x_2^2-x_3^2)\\
			
			4x_1x_2^2x_3(x_2^2-x_1^2-x_3^2) & 4x_1^2x_2x_3(x_1^2-x_2^2-x_3^2) &  -x_3^6-3x_3^2(x_1^2-x_2^2)^2+[3x_3^4+(x_1^2-x_2^2)^2](x_1^2+x_2^2)  
		\end{array}
		\right]
	\end{equation} 
	
	\setcounter{equation}{\value{mytempeqncnt}}
	\hrulefill
	\vspace*{4pt}
\end{figure*}
\setcounter{equation}{30}

\subsubsection{Case II: $1/2<\alpha<1$} 

Following from Eqs.~(\ref{gradient}) and (\ref{Hessian}), we know that neither the monotonicity nor the concavity of the area can be guaranteed when $\alpha>1/2$. Thus, it may be possible to find states and LOCC transformations to violate the LOCC-monotonicity relation.

We consider the W-class state (\ref{W}) with $a^2>1/2$ and $b=c$, and that the local measurement operator $X_k=\begin{bmatrix}
	x_k &0\\
	0 &y_k
\end{bmatrix}$ acts on the state. Then there are $|{\rm W}^{(k)}\rangle=(ay_k|100\rangle+bx_k|010\rangle+bx_k|001\rangle)/\sqrt{p_k}$ with $p_k=|a|^2y_k^2+2|b|^2x_k^2$ for $k=1,2$. Setting $x_1=0$, we have $\mathcal{A}(\ket{{\rm W}^{(1)}})=0$ as $\ket{{\rm W}^{(1)}}$ is a product state. Further, for the W-class state $\ket{{\rm W}^{(2)}}$, there is
\beq
\left(\frac{ay_2}{\sqrt{p_2}}\right)^2=\frac{a^2y_2^2}{a^2y_2^2+2b^2}>\frac{1}{2}
\eeq
when $y_2$ is set to be large enough. Thus, we have $\lambda^A=2b^2/p_2$ and $\lambda^B=\lambda^C=b^2/p_2$ for $\ket{{\rm W}^{(2)}}$. Let $\beta=b^2$ and $g(p_2)=p_2\mathcal{A}(\ket{{\rm W}^{(2)}})$. By simple calculation, we can obtain that the square of $g(p_2)$ can be written as
\beq
g^2(p_2)=\frac{1}{16}p_2^2\mathcal{E}^{2\alpha}(2\beta/p_2)[4\mathcal{E}^{2\alpha}(\beta/p_2)-\mathcal{E}^{2\alpha}(2\beta/p_2)],
\eeq
and its first-order derivative is
\begin{eqnarray}
	&&\frac{{\rm d}}{{\rm d}p_2}g^2(p_2)\nn\\
	&=&2\mathcal{E}^{4\alpha}(2\beta/p_2)\bigg[4p_2\frac{\mathcal{E}^{2\alpha}(\beta/p_2)}{\mathcal{E}^{2\alpha}(2\beta/p_2)}-p_2-8\alpha\frac{\beta\mathcal{E}'(2\beta/p_2)}{\mathcal{E}(2\beta/p_2)}\cdot\nn\\
	&&\frac{\mathcal{E}^{2\alpha}(\beta/p_2)}{\mathcal{E}^{2\alpha}(2\beta/p_2)}+4\alpha\frac{\beta\mathcal{E}'(2\beta/p_2)}{\mathcal{E}(2\beta/p_2)}-4\alpha\frac{\beta\mathcal{E}'(\beta/p_2)}{\mathcal{E}(2\beta/p_2)}\cdot\nn\\
	&&\frac{\mathcal{E}^{2\alpha-1}(\beta/p_2)}{\mathcal{E}^{2\alpha-1}(2\beta/p_2)} \bigg]\nn\\
	&\triangleq& 2\mathcal{E}^{4\alpha}(2\beta/p_2)L(\beta,p_2).\label{derivative}
\end{eqnarray}
Particularly, at the point $p_2=1$, there is
\begin{eqnarray*}
	&&\left.\frac{{\rm d}}{{\rm d}p_2} g^2(p_2)\right|_{p_2=1}\\
	&=&\frac{1}{8}\mathcal{E}^{4\alpha}(2\beta)\bigg[4\frac{\mathcal{E}^{2\alpha}(\beta)}{\mathcal{E}^{2\alpha}(2\beta)}-1-8\alpha\frac{\beta\mathcal{E}'(2\beta)}{\mathcal{E}(2\beta)}\cdot\\
	&&\frac{\mathcal{E}^{2\alpha}(\beta)}{\mathcal{E}^{2\alpha}(2\beta)}+4\alpha\frac{\beta\mathcal{E}'(2\beta)}{\mathcal{E}(2\beta)}-4\alpha\frac{\beta\mathcal{E}'(\beta)}{\mathcal{E}(2\beta)}\frac{\mathcal{E}^{2\alpha-1}(\beta)}{\mathcal{E}^{2\alpha-1}(2\beta)} \bigg]\\
	&\triangleq& \frac{1}{8}\mathcal{E}^{4\alpha}(2\beta) l(\beta).
\end{eqnarray*}
For each measure $\mathcal{E}$ considered in this work, there is
\beq
\lim_{\beta\rightarrow 0} \frac{\mathcal{E}(\beta)}{\mathcal{E}(2\beta)}=\lim_{\beta\rightarrow 0}\frac{\beta\mathcal{E}'(\beta)}{\mathcal{E}(2\beta)}=\lim_{\beta\rightarrow 0}\frac{\beta\mathcal{E}'(2\beta)}{\mathcal{E}(2\beta)}= \frac{1}{2},
\eeq
and thus we can obtain that
\beq
\lim_{\beta\rightarrow 0}l(\beta)=(2\alpha-1)(1-4^{1-\alpha})<0
\eeq
when $1/2<\alpha<1$. It follows from the continuity of $l(\beta)$ that $l(\beta)<0$ for some small $\beta$. Then we know that there must exist $p_{\alpha,\beta}$ such that $[g^2(p_2)]'<0$ for $p_2\in[p_{\alpha,\beta},1]$ (See Fig.~\ref{numerical} (e)-(h) for instances), which implies that $g(p_2)$ is strictly decreasing. Noting that $g(1)=\mathcal{A}(\ket{{\rm W}})$, then we can obtain that 
\begin{equation*}
	\mathcal{A}(\ket{{\rm W}})=g(1)
	<g(p_2)=p_2\mathcal{A}(\ket{{\rm W}^{(2)}}).
\end{equation*}

\subsubsection{Case III: $\alpha=1$}

Here, we construct explicit examples to show that the area with $\alpha=1$ can be increased under certain states and measurements. Consider the pure three-qubit state in standard form of~\cite{Acin2000}
\beq\label{standard}
\ket{\psi}_{ABC}=l_0\ket{000}+l_1e^{i\varphi}\ket{100}+l_2\ket{101}+l_3\ket{110}+l_4\ket{111}
\eeq
with $\sum_{m}l_m^2=1$ and $\varphi\in[0,\pi]$, and the measurement operators are set as $X_i=D_iV$, where
\beq\label{D}
D_1=\begin{bmatrix}
	\sin\varphi_1 & 0\\
	0& \sin\varphi_2
\end{bmatrix},\
D_2=\begin{bmatrix}
	\cos\varphi_1 & 0\\
	0 & \cos\varphi_2
\end{bmatrix},
\eeq
and 
\beq\label{V}
V=\begin{bmatrix}
	\cos\psi_1 & -e^{i\psi_2}\sin\psi_1\\
	\sin\psi_1 &  e^{i\psi_2}\cos\psi_1
\end{bmatrix}
\eeq
with $\varphi_i,\psi_i\in[-\pi,\pi]$. Then through numerical search, we find that the LOCC-monotonicity can be violated by some state. For examples, the area of the triangle enclosed by $\mathcal{W}$ is increased as follows
\beq
\mathcal{A}(\ket{\psi}_{ABC})-\sum_{k=1}p_k\mathcal{A}(\ket{\psi^{(k)}}_{ABC})\approx -0.027,
\eeq
if the state is $l_0\ket{000}+0.264e^{i\varphi}\ket{100}+0.367\ket{101}+0.32\ket{110}+0.055\ket{111}$ with $\varphi=0.8\pi$ and the measurement operators $X_i$ given in Eq.~(\ref{D})-(\ref{V}) are chosen with $\varphi_1=0.4\pi$, $\varphi_2=0.1\pi$, $\psi_1=0.6\pi$, and $\psi_2=0.2\pi$. For the triangle enclosed by $\mathcal{C}^{2\alpha}$, the inequality~(\ref{monotonicity2}) can be violated up to
\beq
\mathcal{A}(\ket{\psi}_{ABC})-\sum_{k=1}p_k\mathcal{A}(\ket{\psi^{(k)}}_{ABC})\approx -0.010
\eeq
when the state is of the form (\ref{standard}) with $l_1=0.096,\ l_2=0.238,\ l_3=0.173,\ l_4=0$, and $ \varphi=0$, and the measurement operators are chosen as Eq.~(\ref{D})-(\ref{V}) with $\varphi_1=0.4\pi,\ \varphi_2=0.2\pi,\ \psi_1=-0.5\pi$, and $\psi_2=-0.1\pi$. For the triangle enclosed by $\mathcal{S}^{\alpha}$, the inequality~(\ref{monotonicity2}) can be violated up to
\beq
\mathcal{A}(\ket{\psi}_{ABC})-\sum_{k=1}p_k\mathcal{A}(\ket{\psi^{(k)}}_{ABC})\approx -0.011
\eeq
when $l_1=0.048,\ l_2=0.046,\ l_3=0,\ l_4=0.141,\ \varphi=0$, $\varphi_1=0.4\pi,\ \varphi_2=0.1\pi,\ \psi_1=0$, and $\psi_2=-0.7\pi$.

\section{Universal triangle relation for pure tripartite states} \label{sec3}

The Tsallis-2 entropy of a quantum state $\rho$ is equal to the state impurity as
\beq
I(\rho):=1-\tr{\rho^2}. \label{impurity}
\eeq
It is always no larger than $1$, as $\tr{\rho^2}\leq 1$ for any physical state $\rho$. If the state is pure, i.e., $\rho=\ket{\psi}\bra{\psi}$ with state vector $\ket{\psi}$, then it has zero impurity. 

Particularly, the state impurity for the discrete quantum system with which the associated Hilbert space has a finite dimension $d$ is given by
\begin{align}
	I(\rho)=1-\sum_{1\leq i \leq d} \lambda^2_i=2\sum_{1\leq i<j\leq d}\lambda_i\lambda_j, \label{qudit}
\end{align}
where $\lambda_i,i=1,\dots, d$ refer to the eigenvalues of $\rho$. For example, any single-qubit state has
\beq
I(\rho)=2\lambda_1\lambda_2=2\lambda(1-\lambda) \label{qubit}
\eeq
with the Schmidt weight $\lambda=\min\{\lambda_1, \lambda_2\}$. It is zero for all pure states and achieves its maximal value $1/2$ with the maximally mixed state $\rho=\id/2$.

For the continuous Gaussian system with infinite state dimension, it is convenient to use the quadrature phase operators $\hat p$ and $\hat q$ to characterize the state. Up to local unitary operations, any $n$-mode Gaussian state can be fully described by a $2n\times 2n$ covariance matrix (CM) $\sigma$, with elements $\an{{{\hat\zeta }_i}{{\hat \zeta }_j} + {{\hat \zeta }_j}{{\hat \zeta }_i}} - \left\langle {{{\hat \zeta }_i}} \right\rangle \left\langle {{{\hat \zeta }_j}} \right\rangle$ where $\an{\bullet}=\tr{\rho\,\bullet}$ and the phase operators are arranged as $\hat{\zeta} = (\hat q_1, \hat p_1, \dots, \hat q_n, \hat p_n)^{\top}$. Correspondingly, its state impurity is determined by the determinant of $\sigma$ as~\cite{Adesso2014}
\beq
I(\rho)=1-\frac{1}{\rt{\det{\sigma}}}. \label{generalmode}
\eeq
When only one single mode is involved, the state impurity further simplifies to 
\beq
I(\rho)=1-1/v.  \label{singlemode}
\eeq
This follows from the fact that the corresponding $\sigma$ has an unique symplectic eigenvalue $v\equiv v_1=v_2$. Generally, the CM as per Eq.~(\ref{generalmode}) becomes insufficient to fully determine the impurity of non-Gaussian states.

The local state impurity of an arbitrary pure bipartite state $\ket{\psi}_{ij}$ is identical to each other, i.e.,
\beq\label{equality}
I(\rho_i)=  I(\rho_j).
\eeq
Here $ I(\rho_{i (j)})$ denotes the impurity of local reduced state $\rho_{i (j)}={\rm Tr}_{j(i)}[\ket{\psi}_{ij}\bra{\psi}]$. In this section, we show that the triangle relation
\beq
\mathcal{E}_{i|jk}\leq \mathcal{E}_{j|ik}+\mathcal{E}_{k|ij} \label{simplified}
\eeq
based on Tsallis-2 entropy or state impurity $I$, is {\it universal} in tripartite quantum systems, in the sense that it is valid for all pure discrete tripartite states, all pure tripartite Gaussian states, and all pure discrete-discrete-continuous states. It is noted that the above relation can be easily generalised to the original one by using the proof of Theorem~1 in the main text.

\subsection{A generic qudit per party} \label{discrete}

For general discrete tripartite states, any pure state can be written in $\ket{\psi}_{ABC}=\sum_{i,j,k}c_{i,j,k}\ket{i_Aj_Bk_C}$ with $\{\ket{i_A}\}$ being a set of state basis for $A$ ($\{\ket{j_B}\}$ for $B$ and $\{\ket{k_C}\}$ for $C$). Each reduced state is generically a qudit, and its state impurity can be given by Eq.~(\ref{qudit}). Note that the state impurity, or equivalently, Tsallis-$2$ entropy, has the subadditivity~\cite{Auden2007,Chehade2019}
\beq
I(\rho_{AB})\leq I(\rho_A)+I(\rho_B) \label{subadditivity}
\eeq
for any discrete bipartite state $\rho_{AB}$. Using the equality $I(\rho_{AB})=I(\rho_C)$ as per Eq.~(\ref{equality}) for any tripartite state $\ket{\psi}_{ABC}$ immediately gives rise to the triangle relation as desired, which has been obtained in the main text.

\subsection{Multi-mode Gaussian state per party} \label{continuous}

Recall first that the CM of any tripartite Gaussian state with mode partition $(n_A, n_B, n_C)$ for $n_i\geq 1$ admits a canonical form of
\beq
\sigma_{ABC}=\begin{pmatrix}
	\sigma_A, \gamma_{AB}, \gamma_{AC} \\ \gamma_{AB}, \sigma_B, \gamma_{BC} \\ \gamma_{AC}, \gamma_{BC}, \sigma_C
\end{pmatrix},
\eeq
where $\sigma_i$ is the CM of reduced $n_i$-mode Gaussian state $\rho_i$ and $\gamma_{jk}$ represents the correlation matrix between party $i$ and $j$. If the state is pure, then it is straightforward to obtain the following equalities
\beq
\det(\sigma_{ABC})=1, \label{ABC}
\eeq 
and
\begin{align}
	\det(\sigma_A)&=\det(\sigma_{BC}),  \\
	\det(\sigma_B)&=\det(\sigma_{AC}),  \\
	\det(\sigma_C)&=\det(\sigma_{AB}),  \label{AB}
\end{align}
by further using Eqs.~(\ref{generalmode}) and~(\ref{equality}). Since $\sigma_{ABC}$ satisfies the strong subadditivity
$\log\det(\sigma_{ABC})+\log\det(\sigma_{C})\leq \log\det(\sigma_{AC}) +\log\det(\sigma_{BC})$~\cite{Lami2016}, substituting it with the above equalities~(\ref{ABC}) to~(\ref{AB}) leads to
\begin{align}
	\det(\sigma_A)&\leq \det(\sigma_B)\det(\sigma_C),  \label{A<BC}\\
	\det(\sigma_B)&\leq \det(\sigma_A)\det(\sigma_C),  \label{B<AC} \\
	\det(\sigma_C)&\leq \det(\sigma_A)\det(\sigma_B).   \label{C<AB}
\end{align}
They are equal to the triangle relation of R\'eny-$2$ entropy of local reduced state which is defined as $R(\rho)\equiv-2\log \tr{\rho^2}=\log\det(\sigma)$ for any Gaussian state $\rho$.

Then, we have
\begin{align}
	&I(\rho_B)+I(\rho_C)-I(\rho_A)\nn \\
	=&1-\frac{1}{\rt{\det(\sigma_B)}}-\frac{1}{\rt{\det(\sigma_C)}}+\frac{1}{\rt{\det(\sigma_A)}} \nn \\
	\geq & 1-\frac{1}{\rt{\det(\sigma_B)}}-\frac{1}{\rt{\det(\sigma_C)}}+\frac{1}{\rt{\det(\sigma_B)\det(\sigma_C)}} \nn \\
	=&\left(1-\frac{1}{\rt{\det(\sigma_B)}}\right)\left(1-\frac{1}{\rt{\det(\sigma_C)}}\right) \nn \\
	=&I(\rho_B) I(\rho_C) \geq 0. \label{proof}
\end{align}
The first equality follows directly from Eq.~(\ref{generalmode}), and the first inequality from Eq.~(\ref{A<BC}). Its permutations under parties $A, B, C$ can be derived similarly, hence proving Eq.~(\ref{simplified}) completely for all pure tripartite Gaussian states.

Finally, given an arbitrary bipartite Gaussian state $\rho_{AB}$, it can be purified to a specific pure tripartite Gaussian state $\ket{\psi_{ABC}}$~\cite{Holevo2001}. In turn, combining the above triangle relation~(\ref{proof}) with Eq.~(\ref{equality}), gives rise to  the impurity subadditivity~(\ref{subadditivity}) and the triangle inequality
\beq
|I(\rho_A)-I(\rho_B)| \leq I(\rho_{AB}) \label{triangle2}
\eeq
for all bipartite Gaussian states. It is worth noting that the above triangle inequality also applies to all discrete bipartite states.

\subsection{The hybrid tripartite state}\label{hybrid}

When the tripartite system is restricted to the discrete-discrete-continuous hybrid, its pure state can be expressed as $\ket{\psi}_{ABC}=\sum_{i,j,k}c_{i,j,k}\ket{i_Aj_B\psi_k}$ with $\{\ket{i_A}\}$ being a set of state basis for $A$ ($\{\ket{j_B}\}$ for $B$) and $\{\ket{\psi_k}\}$ a set of continuous states, including Gaussian and non-Gaussian, for $C$. Generically, the reduced state $\rho_C$ is non-Gaussian. And we are able to obtain
\begin{align}
	I(\rho_C)=I(\rho_{AB})\leq I(\rho_A)+I(\rho_B).
\end{align}
The equality follows from Eq.~(\ref{equality}) and the inequality from the subadditivity~(\ref{subadditivity}) for all discrete bipartite states. Then, assume that $I(\rho_B)\geq I(\rho_A)$, and it is evident that $I(\rho_B)\leq I(\rho_A)+I(\rho_C)$. Additionally, we have 
\begin{align}
	I(\rho_B)+I(\rho_C)-I(\rho_A)&=I(\rho_B)+I(\rho_{AB})-I(\rho_A) \nn \\
	&\geq I(\rho_B)-I(\rho_A)+|I(\rho_B)-I(\rho_A)| \nn \\
	&\geq 0.
\end{align} 
The first inequality derives from the triangle inequality~(\ref{triangle2}) for all discrete bipartite states. Thus, we complete the proof that the triangle relation~(\ref{simplified}) is valid for the pure hybrid tripartite state.

It is remarkable to find that the impurity-triangle relation~(\ref{simplified}) is valid for the general discrete, continuous, and even hybrid tripartite systems. By contrast, the triangle relation, in terms of squared concurrence and Schmidt weight, is only obtained for the $3$-qubit state~\cite{Zhu2015,Qian2018}, and for the discrete tripartite states with Tsallis entropy. Although the R\'enyi-$2$ entropy obeys the triangle relation for both the qubit and the multi-mode Gaussian state~\cite{Lami2016}, it does not apply to qudits~\cite{Linden2013}. It is also interesting to note the equivalence  between the triangle relation as per~(\ref{simplified}) and the subadditivity of entropy or entanglement measure to some extent.

\begin{figure}
	\begin{center}
		\includegraphics[width= 1.0\columnwidth]{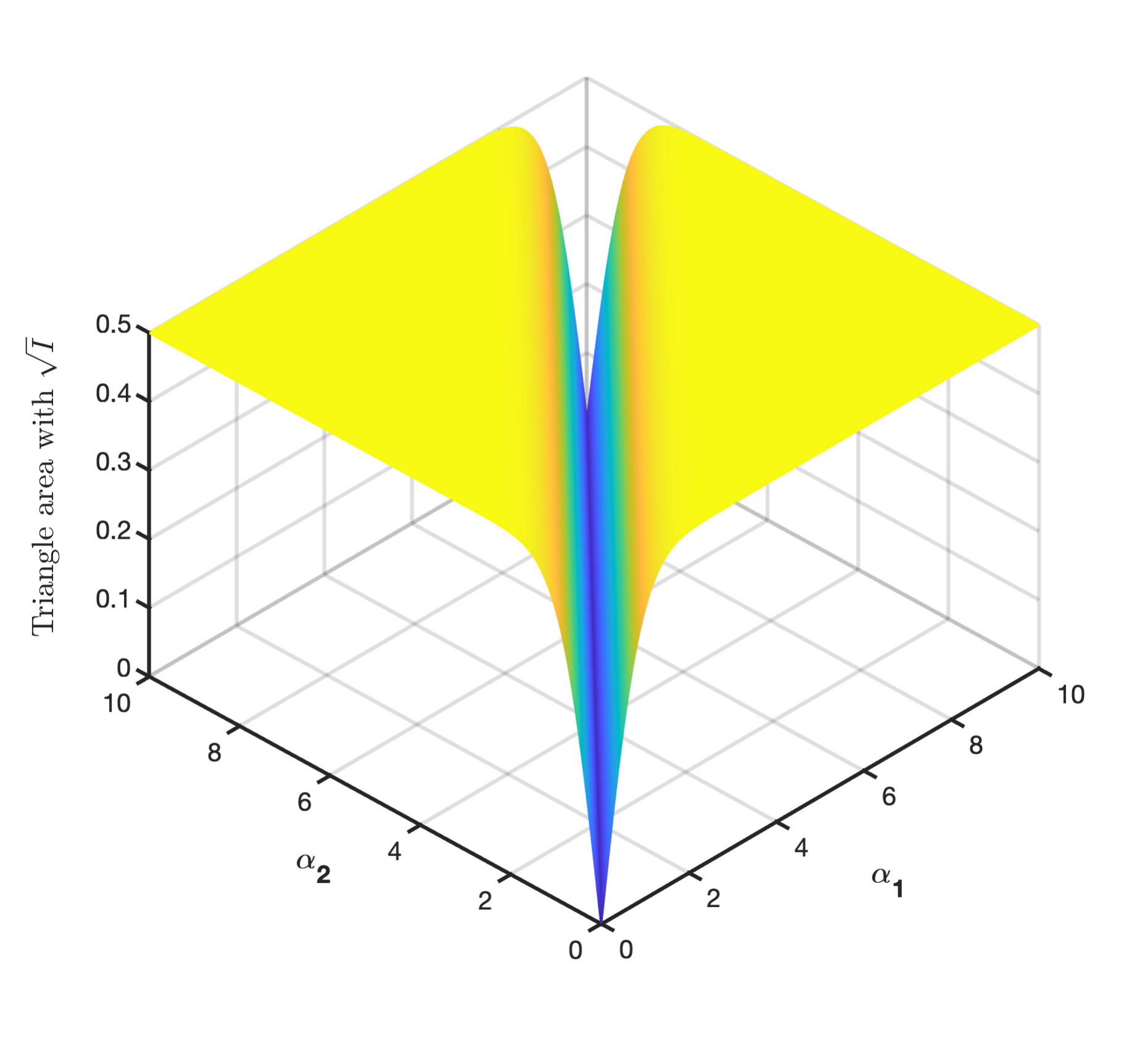}
	\end{center}
	\caption{The triangle area~(\ref{area}) of a class of hybrid states~(\ref{hybridstate}). It is found that the area is almost always nonzero, except for $\alpha_1=\alpha_2$ that the underlying state is bipseparable, and has an upper bound with $\mathcal{A}\leq I(\rho)\leq 1/2$. The normalised coefficient $4/\rt{3}$ has been added to the area.}
	\label{hybridarea}
\end{figure}

\section{Triangle area as a faithful measure for genuine entanglement for discrete, continuous, and hybrid tripartite systems}

Following Proposition~1 and Theorem~1, the triangle relation obtained in the main text provides a geometric picture that is capable of faithfully characterizing and quantifying genuine tripartite entanglement for the discrete, continuous, and even hybrid systems. 

For example, consider a class of hybrid states 
\beq
\ket{\psi}=(\ket{00\alpha_1}+\ket{11\alpha_2})/\rt{2}, \label{hybridstate}
\eeq
where $\ket{\alpha_1}$ and $\ket{\alpha_2}$ are coherent states with real displacements. As plotted in Fig.~\ref{hybridarea}, the area is almost always nonzero, except for $\alpha_1=\alpha_2$ that the underlying state is bipseparable.

We remark that this faithful geometric measure is experimental-friendly, as it requires only the local state information which can be observed within current technology. It is also noted that our work directly adopts the triangle relation to construct feasible measures for genuine tripartite entanglement, whereas previous works~\cite{Adesso2006,Adesso2012,Adesso2014,Li2014,Lami2016} rely on monogamy relations to quantify multipartite Gaussian entanglement. 

\section{Polygon relation for pure multipartite states}\label{sec5}

Given an arbitrary pure $n$-partite state $\ket{\psi}_{A_1\dots A_n}$ that each party $A_i$ is a qudit with dimensionality $d_i\geq 2$ or Gaussian state with modes $n_i\geq 1$, it follows directly from the triangle relation~(\ref{simplified}) that 
\beq
I(\rho_{A_1})\leq I(\rho_{A_2}) +I(\rho_{A_3\dots A_n}).
\eeq
Here the parties $A_3, \dots, A_n$ are grouped into a single party. Dividing the party group $A_3, \dots, A_n$ into the bipartition between $A_3$ and the rest, and then applying the subadditivity of state impurity~(\ref{subadditivity}), yields
\beq
I(\rho_{A_1})\leq I(\rho_{A_2}) +I(\rho_{A_3})+I(\rho_{A_4\dots A_n}).
\eeq
After iterating this bipartition procedure and using the impurity subadditivity with $n-4$ times, we obtain
\beq
I(\rho_{A_1})\leq I(\rho_{A_2}) +I(\rho_{A_3})+\dots+I(\rho_{A_n}). 
\eeq
And its permutations under parties can be easily derived in a similar way. Thus, the local state impurity satisfies a polygon relation for all pure multipartite states, which encompasses the triangle relation for the tripartite state and also significantly generalize the one derived only for the qubit case~\cite{Qian2018}.  We finally remark that following the same argument yields similar polygon relations for any other subadditive measure $\mathcal{E}$, such as von Neumann entropy.

\end{document}